\documentclass[journal=jacsat,manuscript=article]{achemso}

\usepackage[version=3]{mhchem}%
\usepackage{amssymb}
\usepackage{subcaption}
\usepackage{rotating}
\usepackage{braket}
\usepackage{chemfig}
\usepackage{chemformula}
\usepackage{comment}
\usepackage{titlesec}
\usepackage{pgfplots}
\DeclareUnicodeCharacter{2212}{−}
\usepgfplotslibrary{groupplots,dateplot}
\usetikzlibrary{patterns,shapes.arrows}
\pgfplotsset{compat=newest}
\pgfkeys{/pgf/number format/.cd,1000 sep={\,}}

\newcommand*{\E}[1]{\hat{E}_{#1} }

\newcommand*\angstrom{\mbox{\normalfont\AA}} 
\newcommand*\namebond[4][6.5pt]{\chemmove{\path(#2)--(#3)node[midway,sloped,yshift=#1]{#4};}}
\newcommand*\arcbetweennodes[3]{%
    \pgfmathanglebetweenpoints{\pgfpointanchor{#1}{center}}{\pgfpointanchor{#2}{center}}%
    \let#3\pgfmathresult}

\newcommand*\arclabel[6][black,-stealth,shorten <=1pt,shorten >=1pt]{%
    \chemmove{%
        \arcbetweennodes{#4}{#3}\anglestart
        \arcbetweennodes{#4}{#5}\angleend
        \ifdim\anglestart pt>\angleend pt \pgfmathsetmacro\anglestart{\anglestart-360}\fi
        \draw[#1]([shift=(\anglestart:#2)]#4)arc[start angle=\anglestart,end angle=\angleend,radius=#2];%
        \pgfmathsetmacro\anglestart{(\anglestart+\angleend)/2}%
        \node[shift=(\anglestart:#2+1pt)#4,anchor=\anglestart+180,inner sep=1pt,outer sep=1pt]at(#4){#6};%
    }%
}

\usepackage{xr-hyper}
\usepackage{hyperref}
\usepackage{cleveref}
\makeatletter
\newcommand*{\addFileDependency}[1]{
 \typeout{(#1)}
 \@addtofilelist{#1}
 \IfFileExists{#1}{}{\typeout{No file #1.}}
}
\makeatother

\newcommand*{\myexternaldocument}[1]{
 \externaldocument{#1}
 \addFileDependency{#1.tex}
 \addFileDependency{#1.aux}
}

\myexternaldocument{SI}
 
\author{Francesco Mazza}
\affiliation{Scuola Normale Superiore, Piazza dei Cavalieri 7, I-56126 Pisa, Italy}
\author{Marco Trinari}
\affiliation{Scuola Normale Superiore, Piazza dei Cavalieri 7, I-56126 Pisa, Italy}
\author{Chiara Sepali}
\affiliation{Scuola Normale Superiore, Piazza dei Cavalieri 7, I-56126 Pisa, Italy}
\author{Chiara Cappelli}
\affiliation{Scuola Normale Superiore, Piazza dei Cavalieri 7, I-56126 Pisa, Italy}
\email{chiara.cappelli@sns.it}

\title{Analytical Nuclear Gradients for the Multiconfigurational Self-Consistent Field Method Coupled with the Polarizable Fluctuating Charges Model}

\abbreviations{IR,NMR,UV}

\begin{document}

\begin{tocentry}

\includegraphics[]{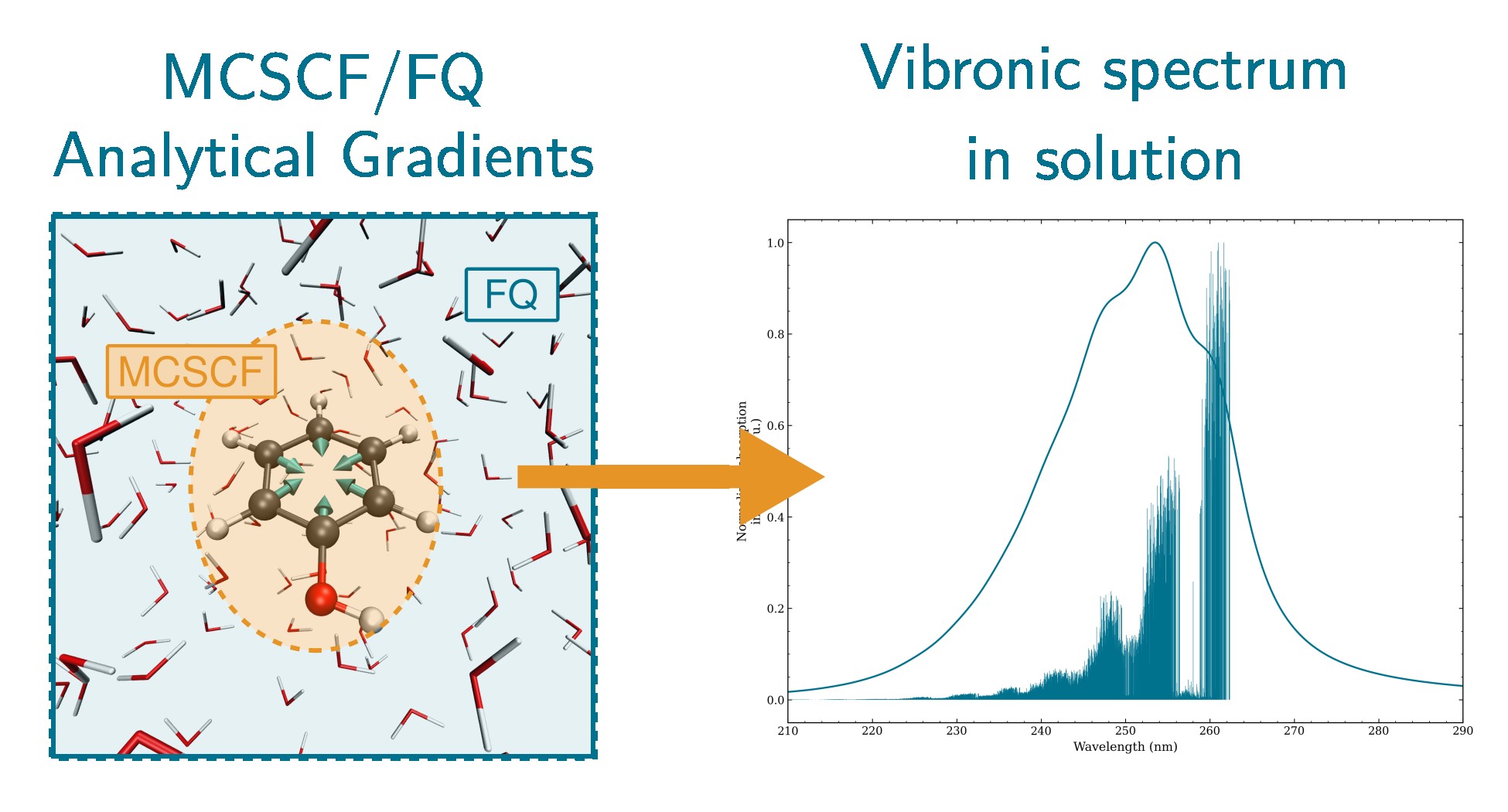}

\end{tocentry}

\begin{abstract}

The multiscale model combining the multiconfigurational self-consistent field (MCSCF) method with the fully atomistic polarizable Fluctuating Charges (FQ) force field (\textit{J. Chem. Theory Comput.} \textbf{2024}, 20, 9954–9967) is here extended to the calculation of analytical nuclear gradients. The gradients are derived from first principles, implemented in the OpenMolcas package, and validated against numerical references. The resulting MCSCF/FQ nuclear gradients are employed to simulate vibronic absorption spectra of aromatic molecules in aqueous solution, namely benzene and phenol. By integrating this approach with molecular dynamics simulations, both solute conformational flexibility and the dynamical aspects of solvation are properly captured. The computed spectra reproduce experimental profiles and relative band intensities with remarkable accuracy, demonstrating the capability of the MCSCF/FQ model to simultaneously describe the multireference character of the solute and its interaction with the solvent environment.

\end{abstract}

\section{Introduction}

There is an ongoing effort in quantum chemistry to investigate molecular systems or processes that exhibit a significant degree of static (or nondynamic) electron correlation. In such cases, the wavefunction cannot be adequately represented by a single electronic configuration, but rather requires the simultaneous consideration of multiple electronic configurations \cite{schmidt1998construction}. These are referred to as multireference systems or processes. Typical examples include transition states, highly conjugated organic molecules (such as polyenes and aromatic systems), and transition metal complexes \cite{keller2015selection}. Processes such as bond dissociation or photoinduced dynamics also fall into this category \cite{choudhury2024understanding, curchod2018ab}. For such cases, qualitatively correct results can only be achieved using multiconfigurational methods. Among these, the multiconfigurational self-consistent field (MCSCF) method \cite{Roos1992multiconfigurational} is one of the most widely employed, as it optimizes both the molecular orbitals and the configuration interaction coefficients simultaneously. 

Despite the high accuracy achievable with multireference methods, they alone cannot adequately describe complex systems subject to numerous intermolecular interactions, such as molecules embedded in condensed-phase environments, including solutions or biological matrices. To model such systems, multireference approaches must be combined with multiscale strategies that partition the system into multiple regions \cite{visscher2011multiscale}. This approach relies on the assumption that only a limited portion of the system is directly responsible for the properties of interest (e.g. a spectral signal), while the surrounding environment mainly acts as a perturbation. Accordingly, the primary region includes the part of the system that gives rise to the property under investigation and is described using a high-level quantum mechanical method. The secondary region, which represents the environment, is modeled at a more approximate level, for instance, by resorting to classical physics.

A widely adopted implementation of this concept is the hybrid quantum mechanics/molecular mechanics (QM/MM) approach, where the primary region is described quantum mechanically and the environment classically \cite{lin2007qm, giovannini2020molecular, senn2009qm}. The classical region can be treated as a dielectric continuum, as in the Polarizable Continuum Model (PCM) \cite{tomasi2005quantum, giovannini2023continuum}, or by retaining an atomistic representation, as in molecular mechanics (MM) \cite{giovannini2020molecular}. While continuum models intrinsically provide a statistical average of the environment and accurately describe long-range electrostatic effects, atomistic models are essential for capturing short-range specific interactions, such as hydrogen bonding \cite{giovannini2023continuum}.  

Among the atomistic polarizable models, \cite{loco2016qm,thole1981molecular,curutchet2009electronic,olsen2011molecular,steindal2011excitation,cappelli2016integrated,giovannini2019polarizable,boulanger2012solvent,rick1994dynamical,thompson1995excited,dziedzic2016tinktep}
the Fluctuating Charges (FQ) approach represents an optimal compromise between accuracy and computational cost. In this model, each atom is endowed with a charge that is not fixed but adjusts to account for the presence of the other charges and the QM molecular potential  \cite{rick1994dynamical, rick1996dynamical, cappelli2016integrated}. The FQ approach belongs to the class of polarizable embedding models and thus accounts for mutual solute (QM) – solvent (FQ) polarization effects.\cite{cappelli2016integrated}
 
For some applications, such as estimating solvation free energies, it may be sufficient to couple the QM and classical regions only at the total energy level. However, for the calculation of molecular properties, it is necessary to extend the formulation to energy derivatives. In particular, the nuclear gradient, which is defined as the first derivative of the energy with respect to the nuclear coordinates, constitutes a key quantity for accessing ground (and excited-state) properties. Due to the scaling of nuclear gradients with the number of atoms (i.e. with the molecule's dimension) analytical formulations are especially desirable, so to enable fast geometry optimizations, the computation of vibronic absorption and emission spectra, and the simulation of nonadiabatic processes in condensed-phase systems \cite{jensen2006introduction, list2013unified, ferrer2012comparison}.  

Based on the aforementioned considerations, this work builds upon the previously developed MCSCF/FQ method \cite{sepali2024fully} implemented in the \textsc{OpenMolcas} package \cite{li2023openmolcas}, extending it to the calculation of analytical nuclear gradients. 

The paper is organized as follows. Section~\ref{sec:theory} provides a brief theoretical overview of the FQ approach, its coupling with an MCSCF wavefunction, together with the derivation of the novel MCSCF/FQ gradient equations. After a brief section focusing on the computational protocols exploited in the study, section~\ref{psec:results} showcases the potentialities of the MCSCF/FQ approach to simulate vibronic absorption spectra of benzene and phenol in aqueous solution, for which experimental spectra have been reported in the literature.\cite{ilan1976photochemistry,riley2018unravelling} A final section summarizes the most relevant findings of the paper and proposes some perspectives for future developments.

\section{Theory}\label{sec:theory}
This section describes the development and implementation of analytical nuclear gradients within the MCSCF/FQ framework. The MCSCF/FQ approach \cite{sepali2024fully} is briefly summarized in Section~\ref{sec:mcscf-fq}, while the derivation of the corresponding analytical gradient equations is presented in Section~\ref{sec:grad}. The theoretical formulation follows the work reported in Ref.~\citenum{sepali2024fully}, adopting the notation introduced by Roos in Ref.~\citenum{Roos1992multiconfigurational} and employing the Einstein summation convention throughout.

\subsection{Multiscale MCSCF/FQ approach} \label{sec:mcscf-fq}

Generally, in a QM/MM multiscale approach, the total energy of the system can be written as a sum of three terms \cite{lin2007qm}:
\begin{equation} \label{eq:eqmemmeqmmm}
  E^{tot} = E^{QM} + E^{MM} + E^{QM/MM}
\end{equation}
where $E^{QM}$ and $E^{MM}$ are the energies of the isolated QM and MM regions, respectively, and $E^{QM/MM}$ represents the interaction term between the two moieties. 
Within the MCSCF/FQ approach\cite{sepali2024fully}, the QM region is described at the MCSCF level. The corresponding energy reads:\cite{Roos1992multiconfigurational}  
\begin{equation} \label{eq:eqnqegy_cas}
  E^{QM} = \braket{\Psi | \hat{H} | \Psi} =   h_{pq} D_{pq} +  g_{pqrs} P_{pqrs} + V^{nn}
\end{equation}
where $h_{pq}$ and $g_{pqrs}$ are the one-electron and two-electron integrals, respectively, and $V^{nn}$ is the nucleus-nucleus repulsion term. ($\boldsymbol{D}$) and ($\boldsymbol{P}$) indicate the first- and second-order reduced density matrices.
The MM region is described using the Fluctuating Charges (FQ) \cite{rick1994dynamical,rick1996dynamical,cappelli2016integrated} polarizable force field. Within this approach, each atom in the classical - FQ - portion is endowed with a charge that can dynamically change to fulfill Sanderson's electronegativity equalization (EE) principle. \cite{sanderson1951interpretation} %
Charges can therefore be obtained through an equivalent reformulation of the EE principle, which involves minimizing the energy functional derived by truncating the Taylor expansion of the energy with respect to the charges up to second order. \cite{rick1996dynamical,cappelli2016integrated} 
To prevent unphysical charge transfer between distant molecules, the total charge $Q_\alpha$ of each molecule $\alpha$ is constrained to remain constant by introducing a set of Lagrange multipliers $\lambda_\alpha$. Therefore, the functional to be minimized, for the isolated FQ region, can be written as \cite{cappelli2016integrated}:
\begin{equation} \label{eq:energy_fq}
  E^{FQ} = \sum_{i\alpha} \chi_{i\alpha}^0 q_{i\alpha} + \frac{1}{2}\sum_{i\alpha, j\beta} q_{i\alpha} T_{i\alpha\, j\beta} q_{j\beta} + \sum_{\alpha} \lambda_\alpha [\sum_i \left(q_{i\alpha}\right) -Q_\alpha ] 
\end{equation}
where $i$ and $j$ run over FQ atoms within each molecule, $\chi_{i\alpha}^0$ is the electronegativity of the isolated atom, and $T_{i\alpha, j\beta}$ is the charge–charge interaction kernel. The diagonal terms of the kernel, $T_{i\alpha, i\alpha}$, account for the contribution $\frac{1}{2}\eta_{i\alpha}q_{i\alpha}^2$ due to the chemical harnesses $\eta_{i\alpha}$. To avoid the so-called ``polarization catastrophe''\cite{thole1981molecular}, the Ohno kernel is used \cite{cappelli2016integrated,ohno1964some}, with the diagonal elements expressed in terms of atomic chemical hardnesses $\eta_{i\alpha}$. The FQ force field thus depends on only two atomic parameters: the electronegativity $\chi^0_{i\alpha}$ and the chemical hardness $\eta_{i\alpha}$. 

In the MCSCF/FQ framework, the interaction energy $E^{QM/MM}$ is the electrostatic interaction between the QM and MM regions. It can be written as: \cite{sepali2024fully}
\begin{equation} \label{eq:eqmmm}
    E^{QM/MM} =  E^{QM/FQ} = q_{i\alpha} V^{i\alpha}_{pq}D_{pq} + q_{i\alpha} V^{i\alpha}_A Z_A
\end{equation}
where $V^{i\alpha}_{pq}D_{pq}$ is the electrostatic potential evaluated at the $i\alpha$-th FQ charge due to the QM electronic density, and  $V^{i\alpha}_A Z_A$ is the analogous potential due to the QM nuclear charges. In this work, a purely state-specific (SS) approach is considered, meaning that both the wavefunction and the FQs are optimized with respect to a single target state. Consequently, the density $D_{pq}$ used to polarize the solvent corresponds to that state.
The interaction kernels  $V^{i\alpha}_{pq}$ and $V^{i\alpha}_A$ are expressed as\cite{sepali2024fully}:
\begin{gather}
    V^{i\alpha}_{pq} = - \Braket{\phi_p | \frac{1}{\left|\boldsymbol{r}_{i\alpha} - \boldsymbol{r}\right|} |\phi_q} \\
    V^{i\alpha}_A = \frac{1}{|\boldsymbol{r}_{i\alpha} - \boldsymbol{R}_A|}
\end{gather}
The total functional to be variationally minimized is therefore \cite{sepali2024fully}:
\begin{multline} \label{eq:ecasfq}
    E(\boldsymbol{D}, \boldsymbol{P}, \boldsymbol{q}, \boldsymbol{\lambda}) = V^{nn} +  h_{pq} D_{pq} +  g_{pqrs} P_{pqrs} + \\
    + \chi_{i\alpha}^0 q_{i\alpha} + \frac{1}{2} q_{i\alpha} T_{i\alpha\, j\beta} q_{j\beta} 
+  \lambda_\alpha [\sum_i \left(q_{i\alpha}\right) -Q_\alpha ] + \\
+ q_{i\alpha} V^{i\alpha}_{pq}D_{pq} + q_{i\alpha} V^{i\alpha}_A Z_A
\end{multline}
The first three terms correspond to the QM energy, obtained as the expectation value of the molecular Hamiltonian (see eq. \ref{eq:eqnqegy_cas}), while the next three terms describe the FQ subsystem (see eq. \ref{eq:energy_fq}). Finally, the last two terms represent the electrostatic interaction between the QM and FQ portions (see eq. \ref{eq:eqmmm}).

Within this scheme, the effect of the FQ charges enters as a perturbation to the MCSCF Hamiltonian, which can be written as follows \cite{sepali2024fully}:
\begin{equation} \label{eq:effectiveH}
  \hat{H}^{eff} =  V^{nn} + q_{i\alpha} V^{i\alpha}_A Z_A + \left[h_{pq} + q_{i\alpha}V^{i\alpha}_{pq}\right]  \hat{E}_{pq} + \frac{1}{2}  g_{pqrs} \left( \E{pq} \E{rs} - \delta_{qr} \E{ps} \right)
\end{equation}
in which the quantity $q_{i\alpha}V^{i\alpha}_{pq}$ is added to the monoelectronic term and $q_{i\alpha} V^{i\alpha}_A Z_A$ to the nuclear repulsion term \cite{sepali2024fully}.

The total MCSCF/FQ energy (eq. \ref{eq:eqmemmeqmmm}) is minimized with respect to all parameters (see Ref.\citenum{sepali2024fully} for more details) by exploiting the usual optimization techniques until self-consistency is achieved. This approach ensures that the mutual polarization between the MCSCF portion and the FQ layer is recovered.

\subsection{Analytical MCSCF/FQ nuclear gradient} \label{sec:grad}

This section introduces a set of novel equations for calculating SS-MCSCF/FQ analytical nuclear gradients. The derivation is given by resorting to a ``fully-focused'' approach. This means that gradients (and related properties) are computed only for the MCSCF portion of the multilayer systems, whereas the explicit terms related to the FQ layer are discarded. Their formulation, taken from Ref.\citenum{lipparini2012analytical}, is recalled in Section \ref{sec:mmgrad} of the Supporting Information (SI). This procedure has been proposed many times in the context of QM/classical approaches and is in line with the so-called Partial Hessian Vibrational Approach (PHVA)\cite{giovannini2019calculation, besley2008partial,vester2024} for vibrational analysis.

The nuclear gradient is the first derivative of the energy with respect to the nuclear coordinates $\xi$:
\begin{equation}
    \frac{d E}{d \xi}  = \frac{d}{d \xi} \braket{\Psi | \hat{H}| \Psi}
\end{equation}
Using the chain rule, this derivative can be expanded as:
\begin{equation} \label{eq:partial_der_energy_general}
     \frac{dE(\boldsymbol{D},\boldsymbol{P},\boldsymbol{q},\boldsymbol{\lambda})}{d\xi} = \frac{\partial E}{\partial \xi} + \frac{d E}{d \boldsymbol{D}}\frac{\partial \boldsymbol{D}}{\partial \xi} +  \frac{d E}{d \boldsymbol{P}}\frac{\partial \boldsymbol{P}}{\partial \xi} +  \frac{d E}{d \boldsymbol{q}}\frac{\partial \boldsymbol{q}}{\partial \xi} +  \frac{d E}{d \boldsymbol{\lambda}}\frac{\partial \boldsymbol{\lambda}}{\partial \xi}
\end{equation}
The MCSCF/FQ energy depends on four sets of parameters: the MOs ($\boldsymbol{\kappa}$), the CI coefficients ($\boldsymbol{C}$), the FQ charges ($\boldsymbol{q}$), and the FQ Lagrangian multipliers ($\boldsymbol{\lambda}$). 
In an SS-MCSCF/FQ calculation, all these parameters are optimized for a specific electronic state, and the corresponding energy derivatives vanish:
\begin{equation} \label{eq:partial_der_zero}
    \frac{d E}{d \kappa} = \frac{d E}{d C} = \frac{d E}{d q} = \frac{d E}{d \lambda} = 0
\end{equation}
This means that the last two contributions to eq. \ref{eq:partial_der_energy_general} vanish.

For an isolated SS-MCSCF system, using the first two conditions in eq. \ref{eq:partial_der_zero} to obtain the derivative of the energy of eq. \ref{eq:eqnqegy_cas} gives: \cite{kato1979energy,schlegel1982mc, yamamoto1996direct, almlof1985molecular}
\begin{equation}
    \frac{d E}{d \xi} = \frac{d V^{nn}}{d \xi} + \frac{d h_{pq}}{d \xi}D_{pq} +  \frac{d g_{pqrs}}{d \xi}P_{pqrs} + F_{pq}\frac{d S_{pq}}{d \xi}
\end{equation}
where the last contribution arises from the derivative of the density matrices with respect to the nuclear coordinates and is related to the orthonormality of the MOs \cite{kato1979energy,schlegel1982mc, yamamoto1996direct, almlof1985molecular}. It can be calculated from the derivative of the overlap matrix $\boldsymbol{S}$ and the generalized Fock matrix $\boldsymbol{F}$ \cite{siegbahn1981complete}:
\begin{equation} \label{eq:generalizeed_fock}
    F_{pq} = D_{pr}h_{qr} + 2 P_{prst} \, g_{qrst}
\end{equation}

Proceeding analogously for the SS-MCSCF/FQ energy, eq. \ref{eq:ecasfq} can be rearranged by grouping terms with the same dependence on the density matrices as follows:
\begin{multline}
        E(\boldsymbol{D}, \boldsymbol{P}, \boldsymbol{q}, \boldsymbol{\lambda}) = \left[ V^{nn} + q_{i\alpha} V^{i\alpha}_{A} Z_A \right] + \left[ h_{pq} + q_{i\alpha}V^{i\alpha}_{pq} \right]D_{pq} + g_{pqrs} P_{pqrs} + \\
         + \chi_{i\alpha}^0 q_{i\alpha} + \frac{1}{2} q_{i\alpha} T_{i\alpha\, j\beta} q_{j\beta} 
+  \lambda_\alpha [\sum_i \left(q_{i\alpha}\right) -Q_\alpha ]
\end{multline}
The last three terms do not contribute to the nuclear gradient because of the conditions shown in eq. \ref{eq:partial_der_zero} and because $\boldsymbol{\chi^0}$, $\boldsymbol{T}$, and $\boldsymbol{Q}$ do not depend on the nuclear coordinates of the QM portion of the system. To evaluate the remaining terms, the same approach as for isolated systems can be followed, resulting in the following equation:
\begin{equation} \label{eq:analytic_grad}
    \frac{d E}{d \xi} = \left[\frac{d V^{nn}}{d \xi} + q_{i\alpha}\frac{d V^{i\alpha}_{A}}{d\xi}Z_A \right] + \left[\frac{d h_{pq}}{d \xi} + q_{i\alpha}\frac{d V^{i\alpha}_{pq}}{d\xi}\right]D_{pq} + \frac{d g_{pqrs}}{d \xi}P_{pqrs}  + \tilde{F}_{pq}\frac{d S_{pq}}{d \xi}
\end{equation}
which includes the FQ contributions. The generalized Fock matrix in the last term is replaced by an effective Fock matrix which accounts for FQ contributions, i.e.:
\begin{equation} \label{eq:effective_fock}
    \tilde{F}_{pq} = D_{pr}\left[h_{qr} +  q_{i\alpha}V^{i\alpha}_{qr} \right] + 2 P_{prst} \, g_{qrst}
\end{equation}
Recalling the QM/MM energy partition of eq. \ref{eq:eqmemmeqmmm}, the gradient contributions in eq. \ref{eq:analytic_grad} can be mapped as follows:
\begin{align}
  \frac{d E^{QM}}{d\xi}\qquad \implies& \qquad \frac{d V^{nn}}{d \xi} + \frac{d h_{pq}}{d \xi}D_{pq} +  \frac{d g_{pqrs}}{d \xi}P_{pqrs} + F_{pq}\frac{d S_{pq}}{d \xi} \nonumber \\ %
    \frac{d E^{QM/MM}}{d\xi} \qquad \implies& \qquad     q_{i\alpha} \frac{d  V_A^{i\alpha}}{d  \xi} Z_A +q_{i\alpha} \frac{d V_{pq}^{i\alpha}}{d \xi} D_{pq} + q_{i\alpha}V^{i\alpha}_{qr}  D_{pr}  \frac{d S_{pq}}{d \xi} \label{eq:eqmmmgrad}  \\
    \frac{d E^{MM}}{d\xi} \qquad \implies& \qquad \varnothing \nonumber
\end{align}

Notice that the approach presented above could in principle be extended to a SA-MCSCF/FQ framework by adopting similar strategies as those developed for alternative polarizable embedding approaches.\cite{song2023state,song2024polarizable, nishimoto2025analytic} However, in that case, since multiple states are included, the evaluation of analytical gradients requires solving coupled-perturbed equations, \cite{staalring2001analytical,osamura1982generalization} which significantly increases the complexity of the derivation. For this reason, in this work, we focus exclusively on SS-MCSCF/FQ gradients.

\subsection{Implementation}

SS-CASSCF/FQ analytical nuclear gradients are implemented in a local version of \textsc{OpenMolcas} \cite{li2023openmolcas}, expanding on prior developments of the CASSCF/FQ model \cite{sepali2024fully}. The implementation is fully integrated into the \emph{Alaska} package \cite{li2023openmolcas} and involves three distinct terms that capture the interaction between the MCSCF and FQ regions, as detailed in \cref{eq:eqmmmgrad}.

The quantities added to the nuclear gradient arise from the interaction energy $E^{QM/MM}$, which in MCSCF/FQ is purely electrostatic. However, non-electrostatic interactions may also play an important role, because they contain the dispersion and the repulsion components.
To recover these contributions, it is possible to resort to a Lennard-Jones potential V(r) \cite{lu2020van}, which for two atoms separated by a distance $r$ reads:
\begin{equation}
    V(r)=4 \varepsilon_{ab} \cdot \left[ \left( \frac{\sigma_{ab}}{r} \right)^{12} - \left(\frac{\sigma_{ab}}{r} \right)^6 \right]
\end{equation}
where the parameters $\varepsilon_{ab}$ and $\sigma_{ab}$ depend on the pair of atom types.
To automatically add this potential to the nuclear gradient, the integration between OpenMolcas and Tinker \cite{rackers2018tinker} is extended to support the CASSCF/FQ model. %

Analytical gradients with respect to MM coordinates have also been developed and implemented (see Section \ref{sec:mmgrad} in the SI).
Notice that, by exploiting this integration, geometry optimizations of the entire system can be performed.

\section{Computational details}

\subsection{Validation step: analytical vs numerical gradients}

\begin{figure}[h!]
     \centering
     \begin{subfigure}[b]{0.30\textwidth}
    \centering
    \includegraphics[width=1.9\textwidth, angle =90]{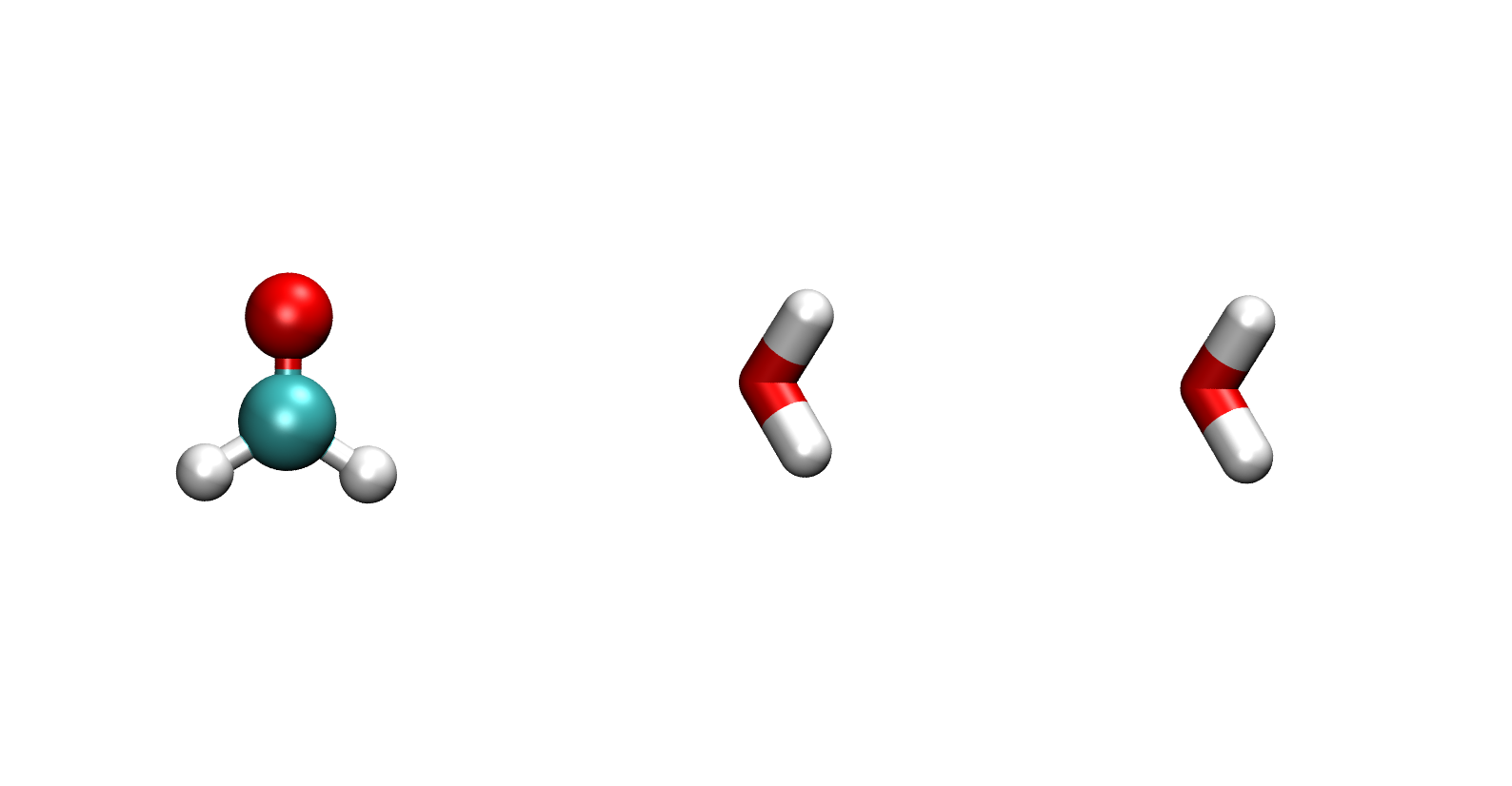}
       \caption{System \textbf{A}\label{fig:sys_a}}
     \end{subfigure}
    \begin{subfigure}[b]{0.30\textwidth}
    \centering
    \includegraphics[width=1.9\textwidth, angle =90]{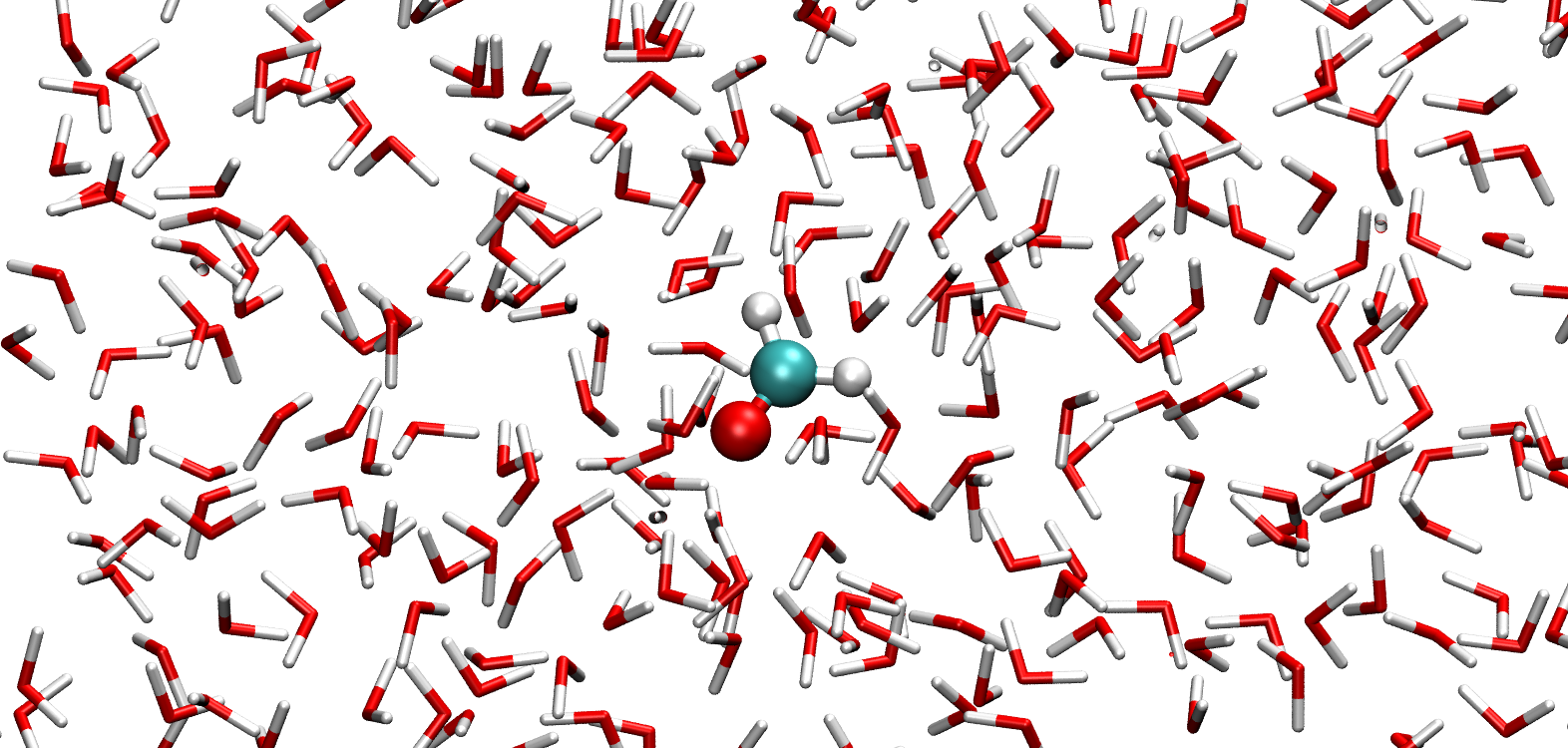}
       \caption{System \textbf{B}\label{fig:sys_b}}
     \end{subfigure}
    \begin{subfigure}[b]{0.30 \textwidth}
    \centering
    \includegraphics[width=1.4\textwidth, angle =90]{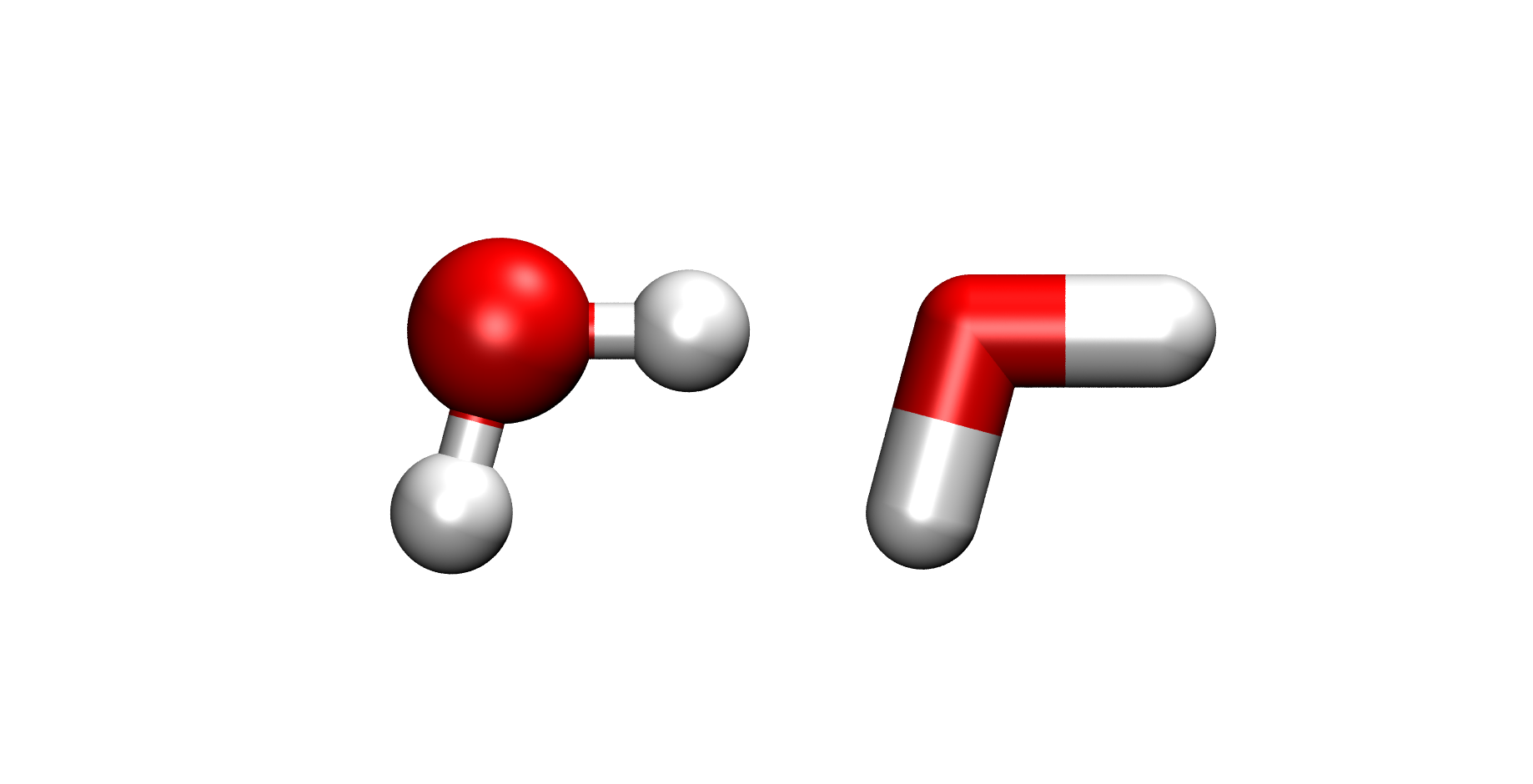}
    \vspace{20pt}
      \caption{System \textbf{C}\label{fig:sys_c}}
     \end{subfigure}

     \caption{Test systems employed for validating and debugging the CASSCF/FQ analytical gradient implementation: a) formaldehyde (QM) and two water molecules (MM) (system \textbf{A}), b) formaldehyde (QM) surrounded by 505 water molecules (MM) (system \textbf{B}), c) two water molecules, one described at the QM level and the other at the MM level (system \textbf{C}).}
     \label{fig:allsystems}
 \end{figure}

The validation of the analytical gradients is performed through comparison with numerical gradients for three model systems (see Figure \ref{fig:allsystems}):
\begin{itemize}
    \item[\textbf{A}.] Formaldehyde (QM) with two water molecules (FQ).
    \item[\textbf{B}.] Formaldehyde (QM) surrounded by 505 water molecules (FQ).
    \item[\textbf{C}.] Two water molecules, one treated as QM and the other as FQ.
\end{itemize} 

 Numerical gradients are computed using the finite difference method up to the eighth order, with a step size of $\Delta = 0.0025\angstrom$ to minimize numerical inaccuracies. Non-equilibrium geometries of the QM molecules are used to amplify gradient components, thereby reducing numerical errors. 
 The non-equilibrium geometry of formaldehyde is obtained by perturbing by a small value ($0.1 \angstrom$) some coordinates of the positions taken from an optimized structure.

For System A, CASSCF/FQ(12,10) gradients are computed using the basis sets CC-PVDZ, CC-PVTZ, aug-cc-PVDZ, and aug-cc-PVTZ. Initial guess orbitals are generated with the GuessOrb program in OpenMolcas \cite{li2023openmolcas}. FQ parameters, referred to as FQ$^a$, are taken from Ref. \citenum{carnimeo2015analytical}. 
For System B, gradients are computed at the CASSCF/FQ(12,10)/CC-PVTZ level of theory, with two sets of FQ parameters being tested: FQ$^a$ \cite{carnimeo2015analytical} and FQ$^b$ \cite{giovannini2019effective}.   
In System C, gradients are evaluated at the CASSCF/FQ$^a$(8,6)/CC-PVDZ level of theory, starting from SCF orbitals. The distance between the QM hydrogen and the FQ oxygen atom is varied from 1.0 to 2.5 \angstrom{} to sample typical hydrogen bond geometries (see Figure \ref{fig:systemC}). As a reference, a single water molecule \textit{in vacuo} is analyzed using SS-CASSCF, which represents an infinite hydrogen bond distance. Only x and y gradient components are considered, as the z-components vanish due to symmetry.

To quantify the agreement between analytical and numerical gradients, three quantities are computed, namely the root mean square deviation (RMSD), the relative root mean square error (RRMSE) and the maximum discrepancy between the components of analytical and numerical gradients ($\Delta_{max}$):
\begin{gather}
    \sigma_a = \operatorname{RMSD} = \sqrt{\frac{1}{3N}\sum_{i=1}^{3N}\left(g_{i}^{analytical}-g_{i}^{numerical}\right)^2} \\
    \sigma_r = \operatorname{RRMSE} = \sqrt{\frac{1}{3N}\sum_{i=1}^{3N}\left(\frac{g_{i}^{analytical}-g_{i}^{numerical}}{g_{i}^{numerical}}\right)^2}
\end{gather}

\begin{equation}
    \Delta_{max} = \max_i(|g_{i}^{analytical} - g_i^{numerical}|)
\end{equation}

\subsection{Vibronic Spectra Calculation}
\begin{figure}[ht!]
  \centering
  \begin{subfigure}{0.49\textwidth}
  \centering
  \includegraphics[width=0.9\textwidth]{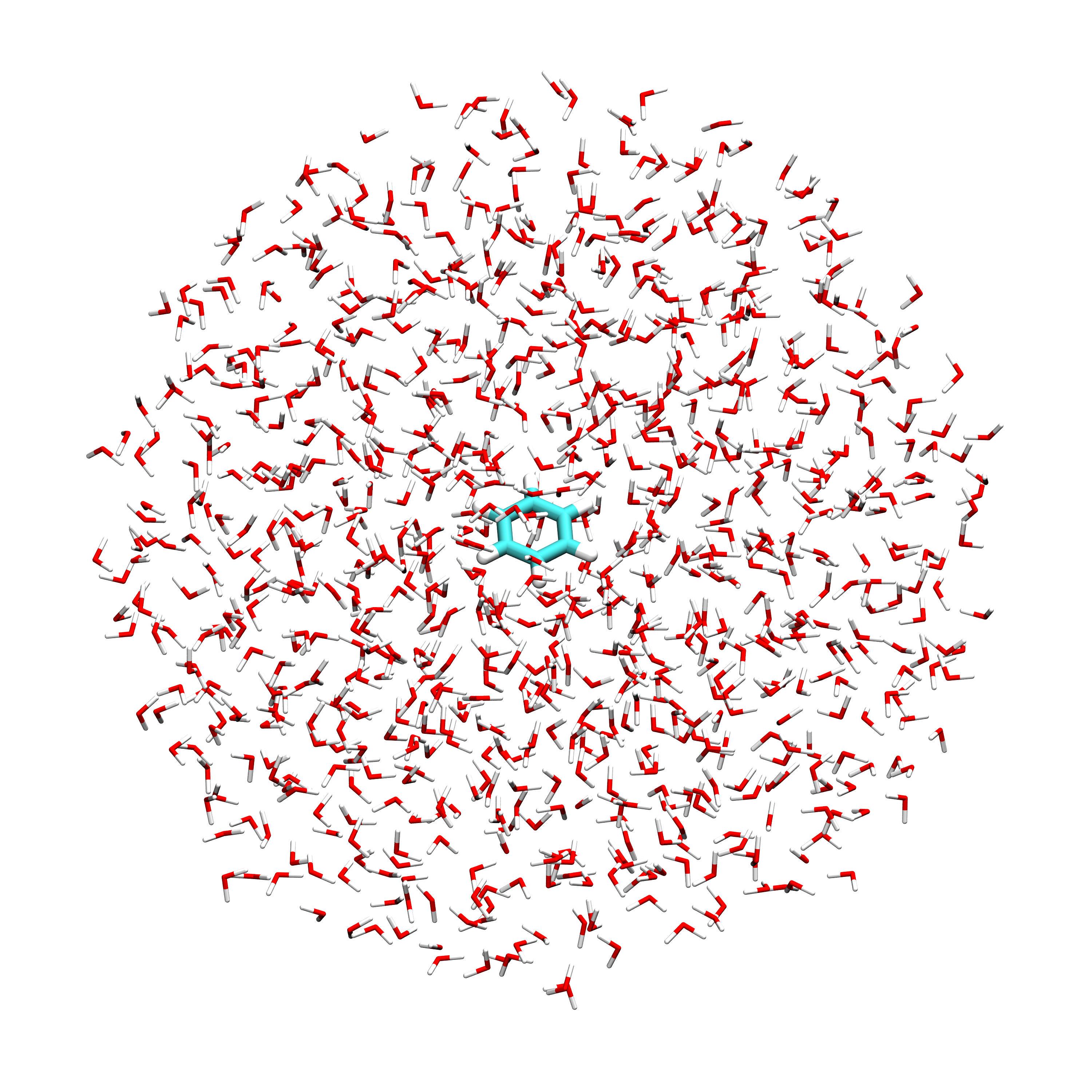}
  \end{subfigure}
  \hfill
  \begin{subfigure}{0.49\textwidth}
  \centering
  \includegraphics[width=0.9\textwidth]{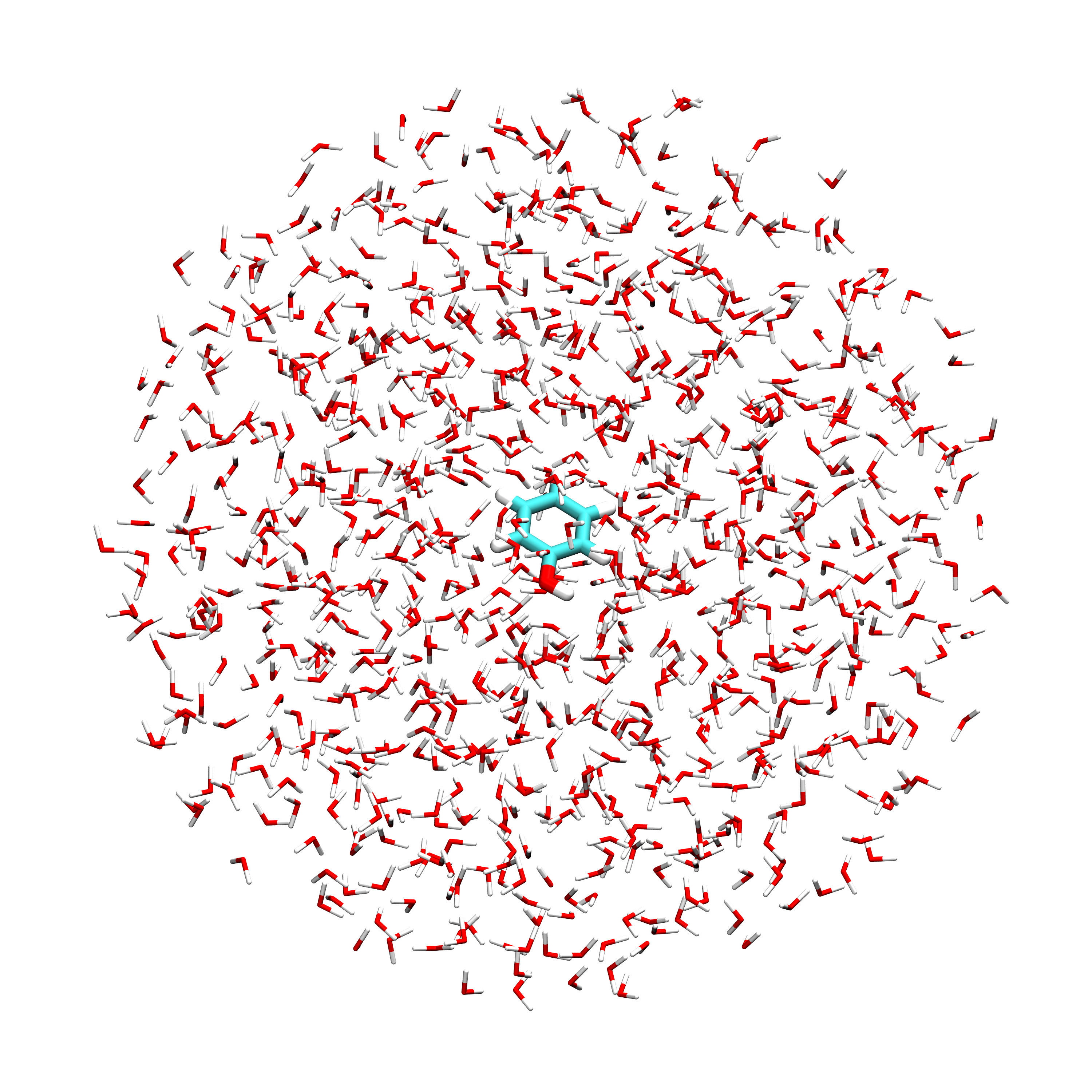}
  \end{subfigure}
  \caption{Representative snapshots of benzene (left) and phenol (right) in aqueous solution, extracted from MD simulations. The radius of the solvation spheres is 18 Å for both solutes and includes $\sim$ 850 water molecules.}
  \label{fig:frame118}
\end{figure}

Vibronic spectra of benzene and phenol in aqueous solution are computed through a multi-step procedure adapted from protocols developed by some of us for simulating molecular spectral signals. \cite{giovannini2020molecular, gomez2022multiple}. The multi-step procedure consists of:\cite{giovannini2020molecular, gomez2022multiple}
\begin{enumerate}
    \item \textit{Definition of the QM/FQ partition:} The solute (QM subsystem) is treated at the CASSCF level, whereas the solvent is modeled with the FQ force field.
    \item \textit{Conformational sampling:} Classical MD simulations are performed over nanosecond timescales to sample the solute-solvent phase space. More details on MD settings for both systems are given in Section \ref{sec:MD} of the SI.
    \item \textit{Extraction of representative structures:} 500 uncorrelated frames are extracted from MD trajectories, and a solvent droplet of sufficient radius is cut around the solute to include long-range interactions between the solute and the solvent. Representative snapshots of benzene (left) and phenol (right) in aqueous solution are shown in Figure \ref{fig:frame118}. 
    \item \textit{CASSCF/FQ calculations:} First, the solutes' geometry is optimized both in the ground (GS) and excited (ES) states. %
    On optimized geometries, SS-CASSCF/FQ calculations are performed to extract all quantities that are required to simulate vibronic spectra (vide infra).%

    \item \textit{Spectra extraction and analysis:} Final vibronic spectra are obtained by averaging computed signals for all snapshots, followed by a convolution with Lorentzian functions. Results are analyzed and compared with experimental data.
\end{enumerate}

The calculation of vibronic spectra is perfomed in the following way. For each of the 500 snapshots, the solute's geometry is first optimized in both the GS and ES with SS-CASSCF/FQ-CC-PVDZ while keeping the FQ solvent molecules frozen, and the Hessians are computed numerically by finite differences of the nuclear gradients using the program \textit{Slapaf} in OpenMolcas\cite{li2023openmolcas}. The ES optimized geometries resulted in structures where the ring was planar, both for benzene and phenol, and this is consistent with previous in vacuo calculations\cite{palmer1993mc, schumm1996casscf}. Then, the geometries are optimized again at the SS-CASSCF/FQ-aug-cc-PVTZ level of theory, and the quantities needed for the subsequent vibronic spectra calculation are computed in both geometries and for both electronic states. These quantities include: energies, nuclear gradients and the transition dipole moment. For aqueous benzene, two sets of FQ parameters are employed and compared, namely FQ$^b$ taken from Ref. \citenum{giovannini2019effective} and FQ$^c$ taken from Ref. \citenum{rick1994dynamical}. For the aqueous solution of phenol only the FQ$^c$ parameters are employed. For benzene, the active space comprises six $\pi$ orbitals with six active electrons, while for phenol it includes seven $\pi$ orbitals with eight active electrons.

Five snapshots are used to perform preliminary tests on CASSCF/FQ parameters, such as the selection of starting orbitals. Specifically, different starting orbitals are tested: Restricted Hartree-Fock (RHF), Unrestricted Hartree-Fock Natural Orbitals (UNO) \cite{keller2015selection, bao2019automatic, toth2016finding}, and Guess orbitals generated by the \textit{GuessOrb} program in OpenMolcas  \cite{li2023openmolcas}. As a result, Guess orbitals are selected as starting orbitals for the benzene solution. For the active space selection of solvated phenol, which presents more challenges, a protocol based on the maximum MO overlap, similar to the one presented by Cárdenas and Nogueira \cite{cardenas2021algorithm}, is used to preserve its consistency among the snapshots, starting from preliminary CASSCF/FQ-CC-PVDZ calculations. The details of this procedure are reported in Section \ref{sec:mosoverlap} in the SI.

Subsequently, for each snapshot, vibronic spectra are calculated with both vertical and adiabatic harmonic approximations\cite{ferrer2012comparison,vibronic_barone}. Vertical approximations describe both the GS and ES PESs by performing the calculation only at the GS optimized geometry. Vertical approximations include Vertical Gradient (VG), \cite{vibronic_barone,macak2000simulations} where the description of the ES PES is obtained with just the ES nuclear gradients, and Vertical Hessian (VH), \cite{hazra2004first} which requires the computation of the ES Hessian. Adiabatic approximations instead describe the ES PES performing the calculations at the ES optimized geometry. Adiabatic approximations include Adiabatic Shift (AS), \cite{vibronic_barone} which describes the ES PES only with the position of its minimum, and Adiabatic Hessian (AH), \cite{vibronic_barone} which requires the computation of the ES Hessian. Among these methods, AS and VG are considerably simpler than AH and VH, because they do not require the calculation of ES Hessian and rely on the approximation that the GS and ES Hessians are equal.
All spectral calculations employ the time-independent (TI) approach, \cite{vibronic_barone} the Frank Condon approximation, at a temperature of 298.15K. For all calculations, it was checked that the recovered fraction of spectra was $\geq 99.9\%$. Spectral calculations are performed using internal coordinates, which have proved to be more robust than Cartesian coordinates, especially employing vertical models \cite{cerezo2016revisiting}.
The final vibronic spectrum is obtained by averaging all spectra of the single snapshots with a convolution with a Lorentzian function with full width at half maximum (FWHM) of 0.040 eV for benzene, and a Lorentzian function with FWHM of 0.15 eV for phenol

The OPLS-AA force field \cite{jorgensen1996development} is used to model Van der Waals interactions in gradient calculations with Tinker \cite{rackers2018tinker}. 
All CASSCF/FQ calculations are performed with a locally modified version of the OpenMolcas software \cite{li2023openmolcas, tange_2024_11247979}, whereas vibronic contributions to electronic spectra are computed with FCclasses3 \cite{cerezo2023fcclasses3}.

\section{Results and Discussion}\label{psec:results}
In this section, we first validate CASSCF/FQ gradients for the model systems depicted in Figure \ref{fig:allsystems}. To this end, beyond checking analytical/numerical consistency, we evaluate the effect of the basis set, FQ parameters, and the size of the water droplet (i.e., the number of water molecules). We further analyze how the distance between the QM and FQ fragments influences CASSCF/FQ nuclear gradients, which also allows for a direct comparison with the analytical–numerical error observed in isolated systems.

Analytical gradients are then employed to simulate vibronic absorption spectra of benzene and phenol in aqueous solution.

\subsection{Analytical gradient validation}

For system \textbf{A}, consisting of formaldehyde (QM) and two water molecules (FQ), numerical and analytical CASSCF/FQ$^a$(12,10) gradients are compared by varying the basis set (CC-PVDZ, CC-PVTZ, aug-cc-PVDZ, and aug-cc-PVTZ). The corresponding results are summarized in Table \ref{tab:sigmaA}. 
Values obtained with or without augmenting the basis set are remarkably similar, thus indicating that augmentation does not improve the results for this simple system.

    \begin{table}[t!]
    \centering
    \begin{tabular}{c|ccc} \\
         Basis set & $\Delta_{max}$ &$\sigma_a$ & $\sigma_r$\\ \hline
         CC-PVDZ & $1.6 \cdot 10^{-6} $ & $6.7 \cdot 10^{-7} $ & $2.3 \cdot 10^{-5}$ \\
         CC-PVTZ &$2.6 \cdot 10^{-6} $ & $1.1 \cdot 10^{-6} $ & $5.8 \cdot 10^{-5}$ \\
         aug-cc-PVDZ &$1.4 \cdot 10^{-6} $ & $5.6 \cdot 10^{-7} $ & $2.7 \cdot 10^{-5}$ \\
         aug-cc-PVTZ &$2.2 \cdot 10^{-6} $ & $1.1 \cdot 10^{-6} $ & $6.1 \cdot 10^{-5}$ 
    \end{tabular}
    \captionof{table}{Maximum deviation between analytical and numerical gradients ($\Delta_{max}$), RMSD ($\sigma_a$), and RRMSE ($\sigma_r$) for system \textbf{A} and different basis sets. The FQ$^a$ paramtrization is employed.  \label{tab:sigmaA}}
    \end{table}

System \textbf{B} (formaldehyde (QM) surrounded by 505 water molecules (FQ)) allows us to assess the impact of both the number of FQ atoms and the choice of the FQ parameterization (FQ$^a$ \cite{carnimeo2015analytical}, FQ$^b$ \cite{giovannini2019effective}) on nuclear gradients. 
CASSCF/FQ$^{(a,b)}$(12,10)-CC-PVTZ results are reported in Table \ref{tab:sigmaB}.%
The comparison with system \textbf{A} values obtained with the same basis set ad parametrization (Table \ref{tab:sigmaA}) shows that increasing the number of water molecules (system \textbf{B}) yields deviations that are about one order of magnitude larger, while the results obtained by changing the FQ parameterization from FQ$^a$ to FQ$^b$ differ by roughly the same scale. This shows that both the size of the solvent droplet and the choice of the FQ parameter sets can influence nuclear gradients to the same extent; nevertheless, deviations remain small, thus indicating that these factors marginally affect the accuracy of analytical gradients with respect to their numerical counterparts.

    \begin{table}[t!]
    \centering
    \begin{tabular}{c|ccc}
         System &$\Delta_{max}$& $\sigma_a$ & $\sigma_r$ \\ \hline
         \textbf{B} (FQ$^a$) &$1.7 \cdot 10^{-5}$& $6.9 \cdot 10^{-6}$& $2.4 \cdot 10^{-4}$ \\
         \textbf{B} (FQ$^b$) &$2.0 \cdot 10^{-6}$& $9.5 \cdot 10^{-7}$& $5.1 \cdot 10^{-5}$
    \end{tabular}
    \captionof{table}{CASSCF/FQ$^{(a,b)}$(12,10)-CC-PVTZ $\Delta_{max}$, $\sigma_a$, and $\sigma_r$ values for system \textbf{B}, as computed by exmploying FQ$^a$ and FQ$^b$ parameter sets. \label{tab:sigmaB}}
    \end{table}

\begin{figure}[t!]
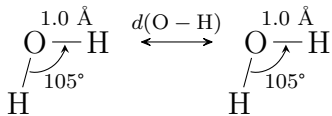

\schemestart 
\setchemfig{atom sep=12pt} \chemfig{[:37.5,2]@{h1}H-[::37.5,2]@{o2}O-[::-75.0,2]@{h3}H}  \namebond{o2}{h3}{\scriptsize 1.0 \angstrom} \arclabel{0.4cm}{h1}{o2}{h3}{\scriptsize105\textdegree}
\hspace{25pt}
 \arrow(@h3--@o5){<->[\scriptsize $d(\ch{O}-\ch{H})$]}
\setchemfig{atom sep=12pt} \chemfig{[:37.5,2]@{h4}H-[::37.5,2]@{o5}O-[::-75.0,2]@{h6}H} \namebond{o5}{h6}{\scriptsize 1.0 \angstrom} \arclabel{0.4cm}{h4}{o5}{h6}{\scriptsize105\textdegree} 
\schemestop
\vspace{37.5pt}
    \caption{System \textbf{C}. One water molecule is treated at the CASSCF level, while the other at the FQ level.}
    \label{fig:systemC}
    \end{figure}

System \textbf{C} is designed to probe the effect of the QM–FQ distance on computed gradients. %
The distance between the QM hydrogen atom and the FQ oxygen atom of the aligned \ch{O–H} bonds (see Figure \ref{fig:systemC}) is varied from 1.0 to 2.5 \angstrom, and extended to infinity (i.e. to a pair of isolated water molecules). The results are summarized in Table \ref{tab:sigmaC}. They show that the QM–FQ distance does not affect the quality of the computed gradients, even in the hydrogen-bonding region. Most importantly, the deviation parameters for the isolated molecule are fully consistent with those obtained using the SS-CASSCF/FQ model, thus demonstrating that the SS-CASSCF/FQ analytical nuclear gradient is as robust as standard CASSCF gradients computed in OpenMolcas.

   \begin{center}
    \begin{tabular}{c|ccc}
         $d(\ch{O}-\ch{H})$ & $\Delta_{max}$ &$\sigma_a$ & $\sigma_r$ \\ \hline
         $1.0\,\angstrom$ &$1.4 \cdot 10^{-7}$ & $8.6 \cdot 10^{-8}$ & $1.0 \cdot 10^{-5}$ \\
         $1.5\,  \angstrom$ &$1.5 \cdot 10^{-7}$ &  $8.7 \cdot 10^{-8}$ &$4.2 \cdot 10^{-6}$ \\
         $2.0\,\angstrom$ &$1.5 \cdot 10^{-7}$ &  $8.7 \cdot 10^{-8}$ & $4.4 \cdot 10^{-6}$ \\
         $2.5\,\angstrom$ &$1.4 \cdot 10^{-7}$ &   $8.6 \cdot 10^{-8}$ & $4.5 \cdot 10^{-6}$ \\
         $\infty$ &$1.4 \cdot 10^{-7}$ & $8.8 \cdot 10^{-8}$ & $4.7 \cdot 10^{-6}$ \\
    \end{tabular}
    \captionof{table}{$\Delta_{max}$, $\sigma_a$, and $\sigma_r$ as a function of the H - O distance in system \textbf{C}. \label{tab:sigmaC}}
       \end{center}%
    
\subsection{Vibronic spectrum of benzene in aqueous solution}

In this section, CASSCF/FQ analytical gradient implementation is applied to simulate the virbonic absorption spectrum of aqueous benzene. Calculations focus on the $S_0 \rightarrow S_1$ electronic transition, and spectra are evaluated using both vertical (VG and VH) and adiabatic (AS and AH) approaches.
  
\begin{figure}
     \centering
     \includegraphics[width=0.9\textwidth]{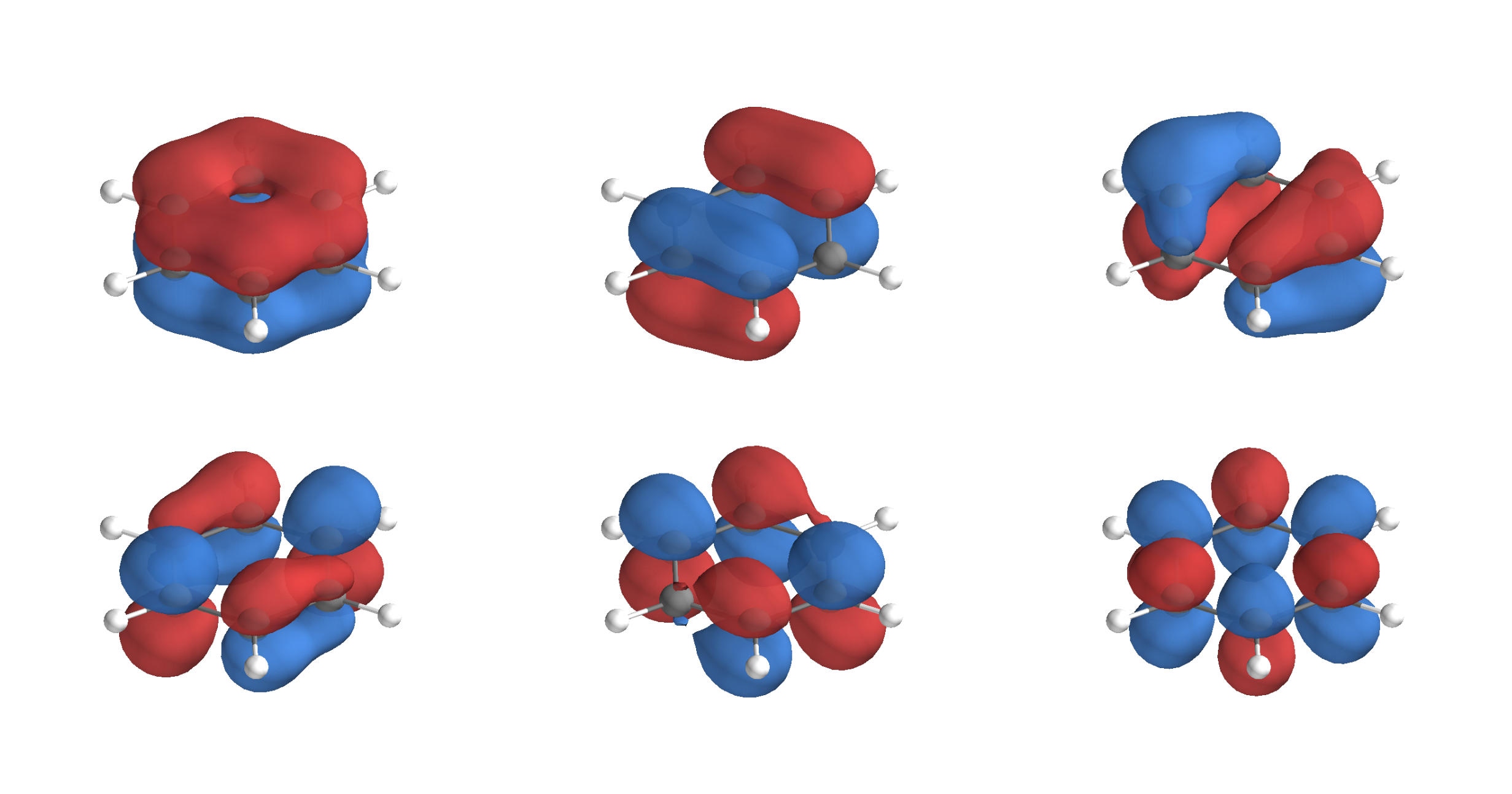}
     \caption{Active orbitals selected for CASSCF(6,6)/FQ calculations. They are obtained with the \textit{GuessOrb} program in OpenMolcas. \cite{li2023openmolcas} }
     \label{fig:active_orb_benzene}
 \end{figure}

Before showing final spectra, preliminary tests are carried out on a limited number of snapshots. %
The simplest way of obtaining the starting MOs for CASSCF calculations consists of performing a preliminary RHF calculation \cite{Roos1992multiconfigurational}. However, RHF orbitals can mix relevant and less relevant contributions, which complicates the identification of the proper active space. Moreover, different snapshots may yield different MOs, hindering the possibility of defining a consistent active space across all geometries. 
Alternative strategies include the use of Natural Orbitals obtained from a UHF calculation (UNO) \cite{keller2015selection, bao2019automatic, toth2016finding} or directly utilizing the guess orbitals generated by the \textit{GuessOrb} program in OpenMolcas \cite{li2023openmolcas}. %
The results obtained by employing all three choices of starting orbitals discussed above are compared for five snapshots. All techniques yield the same converged CASSCF MOs and result in the same energies for benzene. Therefore, guess orbitals are selected in subsequent calculations. %
As already suggested in the literature, \cite{bernhardsson2000theoretical} the active space is select so to include the $6\pi$ orbitals displayed in Figure \ref{fig:active_orb_benzene}.

Another relevant point, which necessitates checking, is the use of an SS-CASSCF approach. When two distinct states, namely the GS and the ES, are obtained through SS-CASSCF calculations, they are optimized independently and, in principle, are neither orthogonal nor non-interacting. %
However, according to the theory, the true states should be two eigenstates of the same Hamiltonian, which are orthogonal to each other. In SS-CASSCF/FQ calculations the Hamiltonians for the GS and the ES are different, because the FQ charges are equilibrated to a specific state of interest (the GS or ES). 
The impact of non-orthogonality on computed excitation energies is evaluated using L\"owdin symmetric orthogonalization, \cite{lowdin1970nonorthogonality, aiken1980lowdin} which minimizes the distance between the original and orthogonalized states. The results show that the correction to the excitation energy is negligible (of the order of $10^{-7}$ a.u. - see Section~\ref{sec:nonorth} in the SI).

\begin{figure}
    \centering
    \includegraphics[width=0.8\linewidth]{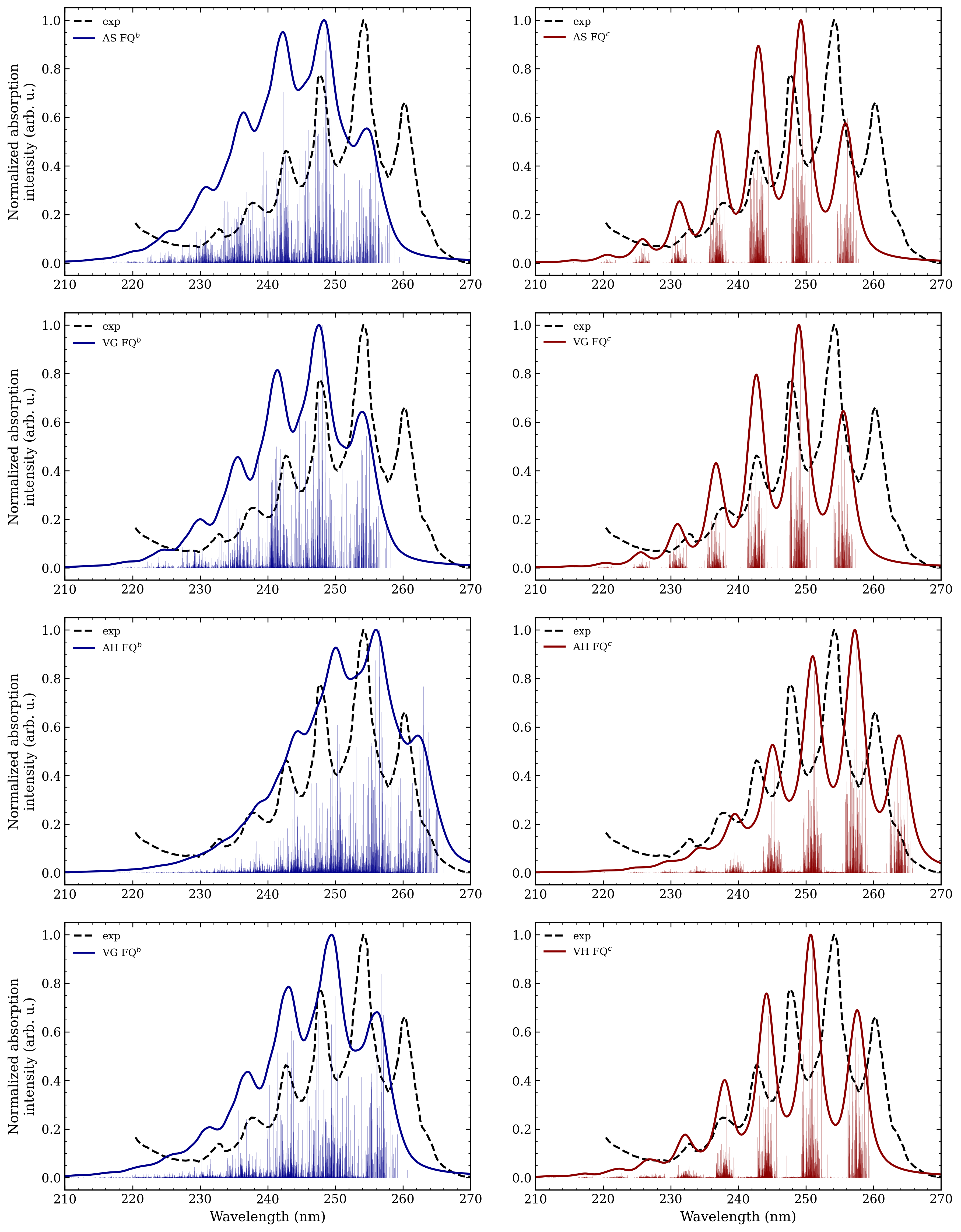}
    \caption{CASSCF(6,6)/FQ$^{(b,c)}$ vibronic spectra of benzene in aqueous solution, obtained with vertical (VG and VH) and adiabatic (AS and AH) approximations. In blue (left), spectra obtained with FQ$^b$ parametrization. In red (right), spectra obtained with FQ$^c$ parameterization. The experimental spectrum (black, dotted) adapted from Ref. \citenum{ilan1976photochemistry} is superimposed.}
    \label{fig:total_benzene}
\end{figure}

Stick and averaged CASSCF/FQ$^{(b,c)}$ vibronic spectra of benzene, obtained from 200 snapshots and computed with vertical (VG and VH) and adiabatic (AS and AH) approximations \cite{ferrer2012comparison}, are reported in Figure~\ref{fig:total_benzene}, together with the experimental spectrum. \cite{ilan1976photochemistry} Sticks represent the set of vibronic transitions computed for each snapshot, with their distribution reflecting the different solvent environments sampled during the MD. In this framework, inhomogeneous broadening arises naturally from conformational sampling, while homogeneous broadening is introduced by convoluting averaged stick spectra with a Lorentzian function with FWHM of 0.04 eV. 
With FQ$^c$ (Figure~\ref{fig:total_benzene}, right), the stick spectra are considerably narrower than with FQ$^b$ (Figure~\ref{fig:total_benzene}, left), leading to sharper convoluted bands. In both parametrizations, however, the AH stick spectra remain significantly broader than those obtained with the other vibronic approximations. Note that 200 snapshots are sufficient to ensure convergence for both FQ parametrizations, as increasing the number of snapshots from 150 to 200 does not alter the results (see Figures~\ref{fig:benz_giov_conv} and~\ref{fig:benz_rick_conv} in the SI).

To evaluate the quality of the convoluted spectra shown in Figure \ref{fig:total_benzene}, three key quantities are considered: the spectral profile, the position of the peaks, and their relative intensities.

The CASSCF/FQ$^{(b,c)}$ spectral profiles exhibit a vibronic structure with clearly distinguishable peaks.
At the CASSCF/FQ$^b$ level of theory (see Figure \ref{fig:total_benzene}, left), VG, VH, and AS approximations yield about six distinct vibronic peaks, while the AH spectrum shows fewer peaks, as excessive broadening prevents a clear resolution of the vibronic structure. Concerning relative intensities, the vertical models predict the three strongest peaks at longer wavelengths, with the central peak being the most intense. In the adiabatic models, instead, the longest-wavelength 0–0 transition is surpassed by the bands at approximately 236 nm (AS) and 244 nm (AH), and therefore does not appear among the three dominant peaks.
Regarding absolute peak positions, the AS, VG, and VH bands appear at comparable wavelengths, with the 0–0 transition around 254 nm (AS and VG) and 256 nm (VH). The AH bands are instead red-shifted by about 8 nm relative to AS/VG and by 6 nm relative to VH.

Moving to CASSCF/FQ$^c$ spectra (Figure \ref{fig:total_benzene}, right), the line shapes are simpler and closely resemble each other, with six distinct and well-separated peaks. The broadening is reduced compared to the CASSCF/FQ$^b$ spectra and remains consistent across all cases, being slightly larger for the AH model. Focusing on relative intensities, the three most intense peaks appear at longer wavelengths also when adiabatic approaches are employed, differently from what we have observed for FQ$^b$ calculations. The peak positions follow the same trend as previously discussed: AS and VG are nearly identical, with the 0–0 transition around 256 nm, while VH gives 258 nm, and the AH spectrum is red-shifted to 264 nm.

We now turn to the comparison with the experimental spectrum. The experimental profile displays a vibronic structure consisting of six main peaks. With the FQ$^c$ parametrization, all experimental bands are well reproduced, and the computed broadening closely matches the experiment. In contrast, FQ$^b$ parametrization yields less resolved peaks, due to a larger broadening. Nevertheless, computed spectra successfully capture the change in slope near the minimum between the two most intense experimental peaks, particularly with VG and VH approaches.

If FQ$^b$ parametrization is employed, AS, VG, and VH show a blue shift of about 4–6 nm for the 0–0 transition (experimentally at $\sim$260 nm), while the AH spectrum is red shifted by about 2 nm. A similar behavior is observed with FQ$^c$, for which AS, VG, and VH give blue shifted values by about 2–4 nm and AH values are red shifted by approximately 4 nm. Overall, the more sophisticated models (AH and VH) provide peak positions in closer agreement with the experimental spectrum. 

Turning to relative intensities, all CASSCF/FQ$^b$ spectra reproduce the experimental feature, where the most intense peak is accompanied by a 0–0 transition at roughly 60\% of its intensity. Differences emerge in the secondary peaks: in the adiabatic models (AS and AH), the peak to the left of the most intense one (236 nm for AS and 244 nm for AH) is systematically overestimated, whereas in the vertical models (VG and VH) the intensities show very good agreement with experiment. The experimental ordering of intensities is preserved in all cases when FQ$^c$ is employed, with VG and VH providing the closest match, while AS and AH tend to slightly overestimate the peaks to the left and underestimate the 0–0 band. 
The difference between the spectra obtained with the two FQ parametrizations is consistent with their intrinsic characteristics. The broader profiles associated with FQ$^b$ likely stem from the fact that FQ$^c$ parameters are optimized to reproduce the properties of bulk water \cite{rick1994dynamical}, whereas FQ$^b$ parameters are explicitly designed to describe electrostatic and polarization interactions between solutes and the aqueous solvent \cite{giovannini2019effective}. Nonetheless, both parametrizations, combined with the different vibronic approaches, can capture the main experimental features. Overall, the VG and VH spectra provide a more accurate reproduction of the experimental profile than AS and AH. This discrepancy can be rationalized by noting that, while adiabatic models offer a more accurate description of the final PES, vertical approaches better represent the vertical region of the PES, which dominates the most intense peak \cite{bloino2016aiming}.

\subsection{Vibronic spectrum of phenol in aqueous solution}

We now move to a more challenging case: aqueous phenol. Unlike benzene, phenol features an explicit site for hydrogen bonding with the surrounding water, making an accurate treatment of the explicit solvent around the solute essential. In fact, it is reasonable to assume that, in the case of phenol, a reliable reproduction of the vibronic spectrum requires a subtle interplay among conformational analysis, the characterization of the solvent arrangement around the solute, correlation effects, and the level of treatment of vibronic couplings. 

\begin{figure}
    \centering
    \includegraphics[width=0.9\linewidth]{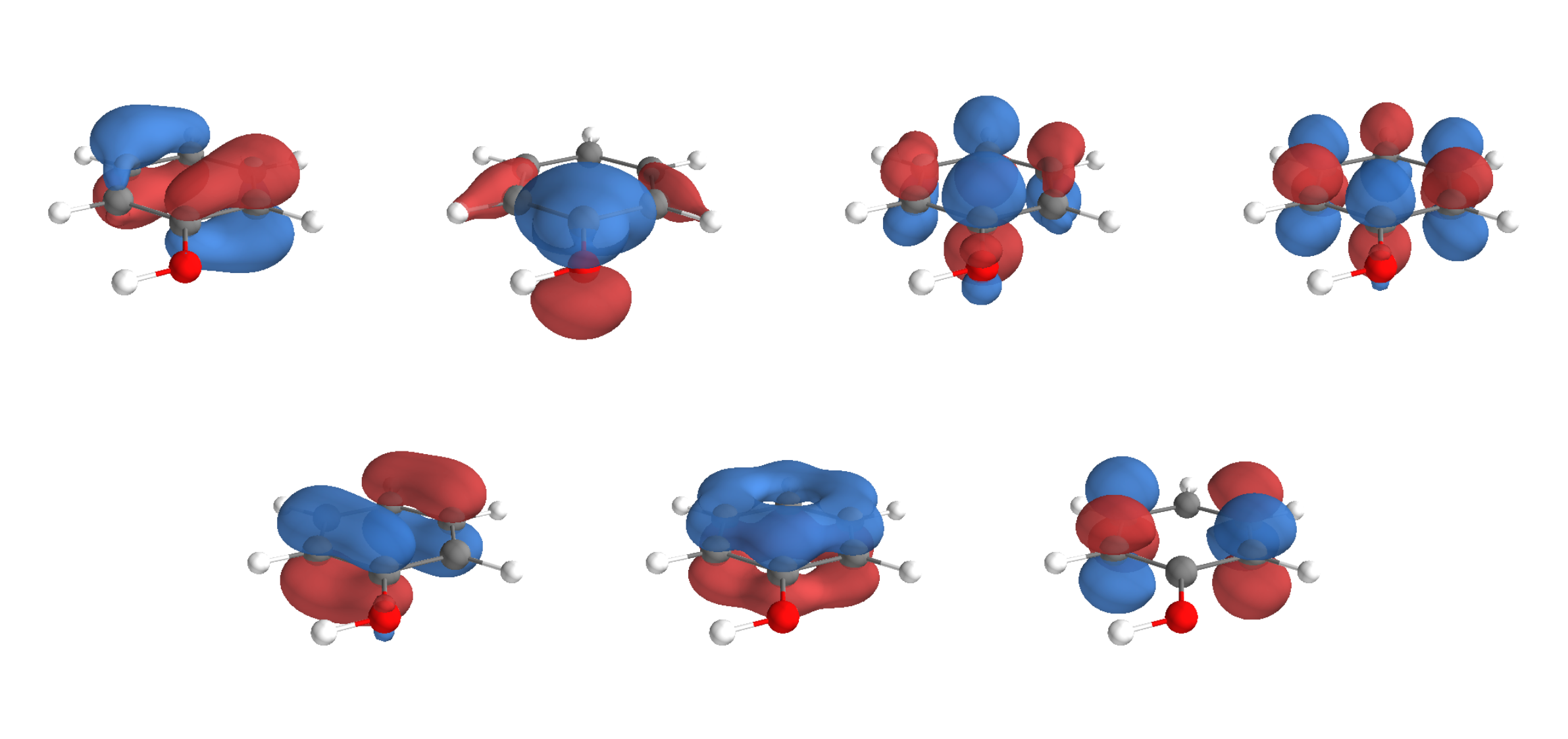}
    \caption{Active orbitals selected for CASSCF(8,7)/FQ calculations of phenol in aqueous solution. This active space is composed of the 7 phenol $\pi$ orbitals, in which the orbital localized mainly on the oxygen can appear distorted as a consequence of the OH bond out-of-plane rotation.}
    \label{fig:active_orb_phenol}
\end{figure}

Similar to benzene, VG, VH, AS, and AH approaches are challenged, coupled with CASSCF(8,7)/FQ$^c$-aug-cc-PVTZ calculations. The employed active space includes the seven $\pi$ orbitals of phenol, in agreement with previous studies\cite{schumm1996casscf}. The presence of the OH group introduces additional complexity, since during the classical MD (step 2 of the computational protocol reported in Section \ref{psec:results}) the OH bond may rotate out of the ring plane. This rotation complicates the definition of the active space compared to benzene. For most snapshots, the selected active orbitals correspond to those reported in Figure \ref{fig:active_orb_phenol}. To generate suitable starting orbitals, RHF orbitals, UNO, GuessOrb-generated orbitals, and preliminary CASSCF(8,7)/FQ$^c$ calculations are compared. Only the latter consistently provides a comparable set of MOs across all snapshots and is therefore adopted in the subsequent calculations. Consistency in the choice of the active space among different sets is ensured through the protocol detailed in Section \ref{sec:mosoverlap} in the SI.

\begin{figure}
    \centering
    \includegraphics[width=0.8\linewidth]{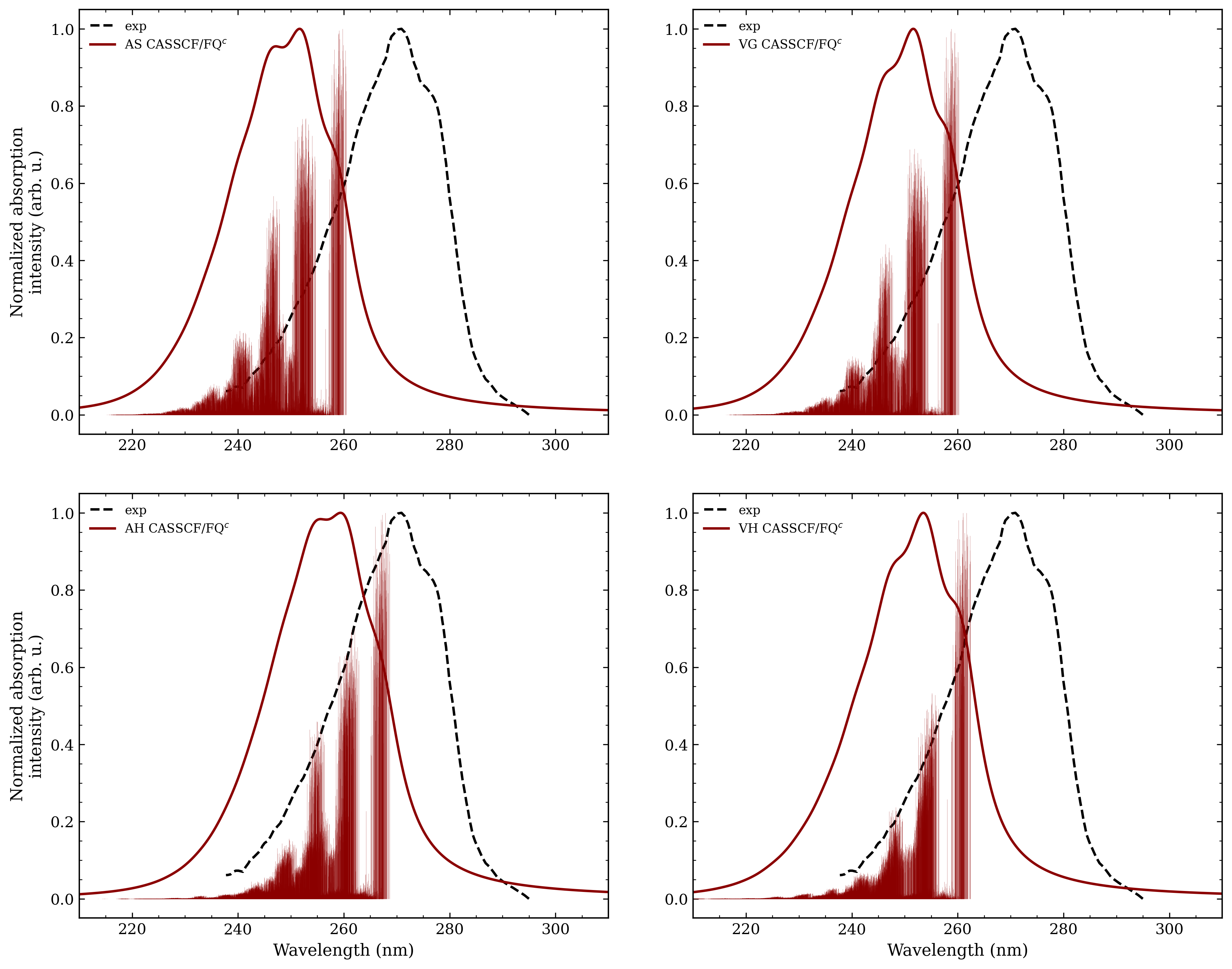}
    \caption{Stick and convoluted CASSCF(8,7)/FQ$^{c}$ vibronic spectra of phenol in aqueous solution, obtained using the vertical (VG and VH) and adiabatic (AS and AH) approaches. The experimental spectrum (black, dotted) adapted from Ref. \citenum{riley2018unravelling} is superimposed.}
    \label{fig:total_phenol}
\end{figure}

In Figure \ref{fig:total_phenol}, CASSCF/FQ$^c$ vibronic spectra are shown for all four harmonic approximations, including both convoluted spectra and stick transitions. The experimental spectrum \cite{riley2018unravelling} is also reported for the sake of comparison. Similar to benzene, the variability of the vibronic signals across 150 snapshots accounts for the inhomogeneous broadening, while the Lorentzian convolution with a FWHM of 0.15 eV recovers the homogeneous broadening. Notice that 150 snapshots are enough to reach convergence, as shown in Figure~\ref{fig:phenol_conv} in the SI. 
Stick spectra show a vibronic structure with three main peaks. At longer wavelengths, peaks appear more isolated, while in the central region, a denser distribution of signals is observed. This leads to broad convoluted spectra, of which the intensity at higher wavelengths decreases steeply. All vibronic approaches provide a similar shape for the rightmost region, but AS and AH display higher intensities for the other features compared to the VG and VH, leading to clear differences in the overall spectral shape.

Convoluted spectra display a broad main band with two vibronic shoulders (see Figure \ref{fig:total_phenol}). The central peak is the most intense across all vibronic models, while the red-most peak is the weakest. However, the relative intensities differ between vertical and adiabatic approaches: unlike vertical models, adiabatic approaches predict the low-wavelength peak to be nearly as intense as the central band. The positions of the peaks in AS and VG spectra are almost perfectly aligned, with the main peak at about 250 nm. The VH spectrum is slightly blue-shifted (main peak at 253 nm), while the AH spectrum shows a more pronounced blue shift of about 6 nm with respect to vertical approaches. Note that a similar trend was observed in the previous section for benzene in aqueous solution.
The experimental spectrum (see Figure \ref{fig:total_phenol}) presents a single broad peak that features two clearly visible vibronic shoulders, one on the left and one on the right at the top of the peak. Overall, these features are well reproduced by calculations. In both the computed and experimental VG and VH spectra, the shoulders reach about 80\% of the main peak intensity. AS and AH overestimate the intensity of the left shoulder, and this results in a line shape that is less in accordance with the experimental profile.
The position of the main peak presents a large blue shift of about 15 to 20 nm for the AS, VG and VH spectra and about 10 nm for the AH spectrum. These shifts, which are much larger than those observed for benzene, are probably caused by the explicit hydrogen bonding interaction between phenol and benzene, which requires an accurate description of both the flexibility of the phenolic OH group, and the specific arrangement of water around it. Another concurrent factor that can explain this larger shift can be the lack of dynamical correlation in CASSCF calculations.

In conclusion, as in the case of the benzene aqueous solution, vertical approaches produced spectra more in accordance with the experimental absorption spectrum than the adiabatic approximations, with little differences between the VG and the VH models. This suggests that, also in this case, an accurate description of the PES around the Franck–Condon region, as achieved with the vertical models, is more critical than an overall accurate representation of the excited-state PES obtained with the adiabatic approximations. \cite{bloino2016aiming}

\section{Summary, Conclusions and Future Perspectives}

We have presented the development and implementation of multiscale SS-MCSCF/FQ analytical nuclear gradients. The model has been validated through comparison between analytical and numerical gradients.
Subsequently, analytical CASSCF/FQ nuclear gradients were employed to simulate the vibronic spectra of aqueous benzene and phenol, using four different approximations to describe vibronic progressions \cite{ferrer2012comparison,bloino2016aiming}.

All tested vibronic schemes, combined with the various FQ parameterizations, successfully reproduce the main experimental features, including spectral profiles, peak positions, and relative intensities. Overall, vertical approximations yield spectra in closer agreement with experiments. In particular, the VG model proved nearly as effective as the more elaborate VH approach, providing a computationally cheaper yet reliable alternative.

The good agreement between computed and experimental spectra also confirms that the CASSCF/FQ approach effectively captures both the multireference character of benzene and phenol and their interactions with the aqueous environment.

Future developments will focus on achieving quantitatively accurate results by including dynamical correlation effects. Starting from a multiconfiguration reference, these could be incorporated via a perturbative treatment \cite{andersson1990second}, for instance, through the development of a CASPT2/FQ scheme.\cite{nishimoto2025analytic}

Moreover, the generality of the present formulation makes it suitable for extension to treat other solvents\cite{ambrosetti2021quantum} or more complex environments.\cite{gomez2023uv} Another important perspective concerns the inclusion of non-electrostatic solute–solvent interactions, which are not accounted for in the current electrostatic-only FQ model. In particular, quantum repulsion and other short-range effects are expected to play a significant role and will be essential to achieve quantitatively accurate descriptions of solvation.\cite{giovannini2019quantum,giovannini2017general, jensen1998approximate, amovilli1997self} 

Finally, the extension of the analytical nuclear gradient to a state-average formalism \cite{nishimoto2025analytic} would open the way to the study of photochemical processes \cite{choudhury2024understanding, curchod2018ab} involving multiple excited states\cite{ruckenbauer2016revealing, crespo2018recent, lischka2018multireference, plasser2019highly,cecam}.

\begin{acknowledgement}

The authors acknowledge funding from MUR-FARE Ricerca in Italia: Framework per l'attrazione ed il rafforzamento delle eccellenze per la Ricerca in Italia - III edizione. Prot. R20YTA2BKZ and the European Union's Horizon Europe research and innovation programme under the project HORIZON-MSCA-2023-DN-01 - LUMIÈRE G.A . No 101169312. The Center for High-Performance Computing (CHPC) at SNS is also acknowledged for providing the computational infrastructure.

\end{acknowledgement}

\begin{suppinfo}

Nuclear gradients with respect to MM coordinates. Computational details and analysis of MD simulations of benzene and phenol in aqueous solution. An algorithm to select active spaces among different structures of the same system, using MO overlap. Estimates of the effects of Non-Orthonormality of SS wavefunctions. Convergence tests on the number of snapshots used for computing vibronic spectra.

\end{suppinfo}

\bibliography{bibliografia_abstract_tesi}

\end{document}


\section{Nuclear gradients with respect to MM coordinates} \label{sec:mmgrad}

The derivative of the energy with respect to the position of MM atoms ($\xi_{MM}$) arises from $E^{QM/MM}$ and $E^{MM}$ energy contributions (see Equation 1 in the main text), because $E^{QM}$ does not depend directly on MM coordinates. Using the chain rule and the variational conditions, the nuclear gradient can be expressed as:
\begin{equation}
    \frac{dE}{d\xi_{MM}} = \frac{dE^{QM/MM}}{d\xi_{MM}} + \frac{dE^{MM}}{d\xi_{MM}}  = \sum_{i\alpha} q_{i\alpha} \frac{d V_{i\alpha}(D)}{d\xi_{MM}} +  \frac{1}{2} \sum_{i\alpha, j\beta} q_{i\alpha}\frac{dT_{i\alpha , j\beta}}{d\xi_{MM}} q_{j\beta}
\end{equation}

where the derivative of the interaction energy $E^{QM/MM}$ is the product between the FQ charges and the electric field, generated by the QM density, that acts on them. The only contribution that arises from the MM energy is due to the derivative of the charge-charge interaction kernel. In case the Onho kernel \cite{cappelli2016integrated,ohno1964some} is employed, the derivative of the kernel can be written as:
\begin{equation}
    \frac{d T_{i,j}}{dr_i} = -\frac{\eta_{ij}^3}{(1+\eta_{ij}^2 r_{ij}^2)^{\frac{3}{2}}} r_{ij}
\end{equation}
where in this case, $i$ and $j$ are FQ atoms, $r_i$ indicates the position of atom $i$, $r_{ij} = r_i - r_j $, and $\eta_{ij}$ is the arithmetic mean between the chemical hardnesses of atom $i$ and atom $j$.

\section{MD simulations}
\label{sec:MD}

Molecular dynamics (MD) simulations are carried out for both benzene and phenol in aqueous solution, using the General Amber Force Field (GAFF) \cite{gaff}. GAFF parameters are generated with Antechamber \cite{wang2006automatic} for both solutes, while charges are derived from RESP scheme for benzene \cite{bayly1993well} and from CM5 model for phenol \cite{marenich2012charge}.
Simulations are performed with the GROMACS package \cite{Lindahl15_19}. The solvation box is a cube with sides of 5.5 nm, containing about 5300 TIP3P \cite{Nilsson01_9954} water molecules and one solute molecule in the case of benzene, and about 5600 water molecules in the case of phenol. The steepest descent minimization algorithm is used to minimize energy during the structure's initial relaxation of both solutes. Then, two equilibration steps are carried out. For benzene, an equilibration (0.5 ns) under the NVT ensemble is performed to reach a temperature of 298 K using a coupling constant of 0.1 ps and a velocity-rescaling method \cite{bussi2007}, followed by 1 ns of NPT equilibration with the Parrinello–Rahman barostat \cite{parrinello1982}. For phenol, the corresponding equilibration times are 1 ns (NVT) and 2 ns (NPT). The MD production stage lasts 30 ns in both cases, using the leap-frog integration algorithm \cite{berendsen1986practical} with a time step of 1 fs for benzene and 2 fs for phenol.
Electrostatic interactions are described with the Particle Mesh Ewald method \cite{darden1993pme}. %

The Radial Distribution Function (RDF) and Dihedral Distribution Function (DDF) for phenol are plotted in Figure S1. The RDF shows that a hydrogen bond is formed between the phenolic hydrogen (H6) and the oxygen atoms of surrounding water molecules, with a pronounced peak centered at 2 Å. The corresponding running coordination number (RCN) of approximately 1 indicates that, on average, one hydrogen bond is formed. In addition, the C2–C1–O1–H6 dihedral angle, denoted as $\delta_1$, is analyzed along the MD trajectory, showing the expected rotational flexibility of phenolic hydrogen H6, with peaks centered at 0° and ±180°. For comparison, the RDF of benzene is reported in Figure S2, clearly indicating the absence of hydrogen bonding between the benzene hydrogens and the oxygen atoms of water molecules.

    \begin{figure}[!ht]
    \centering
    \includegraphics[width=0.8\linewidth]{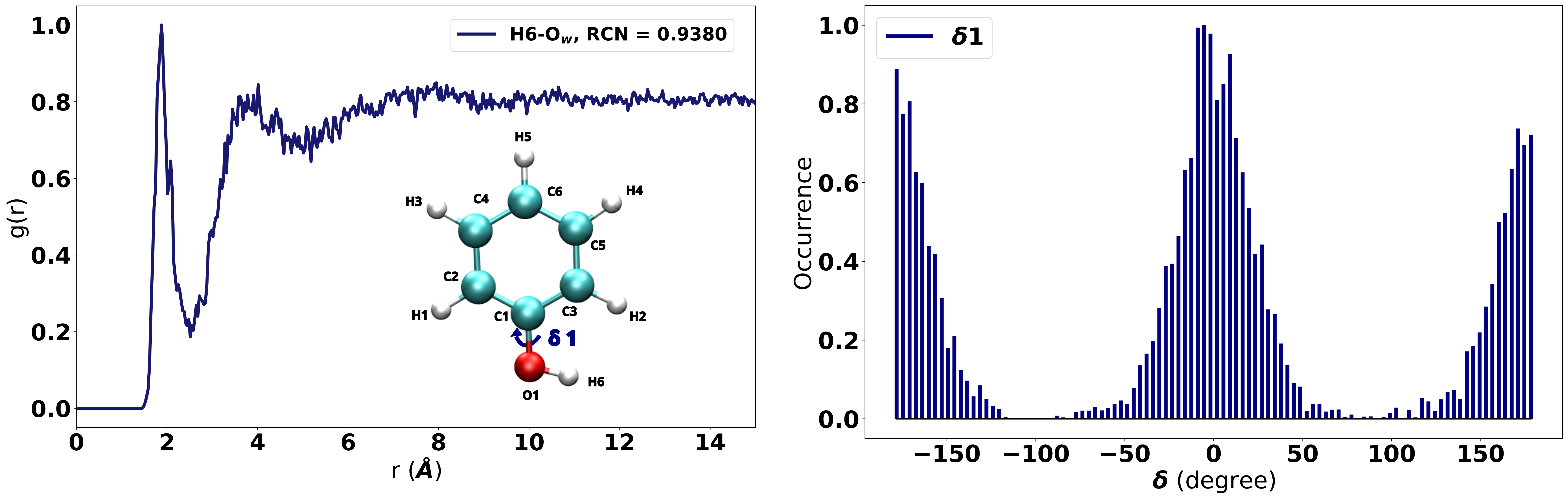}
    \caption{a) Left: Definition of  Phenol dihedral angle $\delta_1$,  and its corresponding DDF.
    (b) Right: RDF between the phenol hydrogen (H6) and water oxygen atoms, along with the corresponding running coordination number (RCN).}
    \label{f:rdf_ddf_phenol}
    \end{figure}

    \begin{figure}[!ht]
    \centering
    \includegraphics[width=0.8\linewidth]{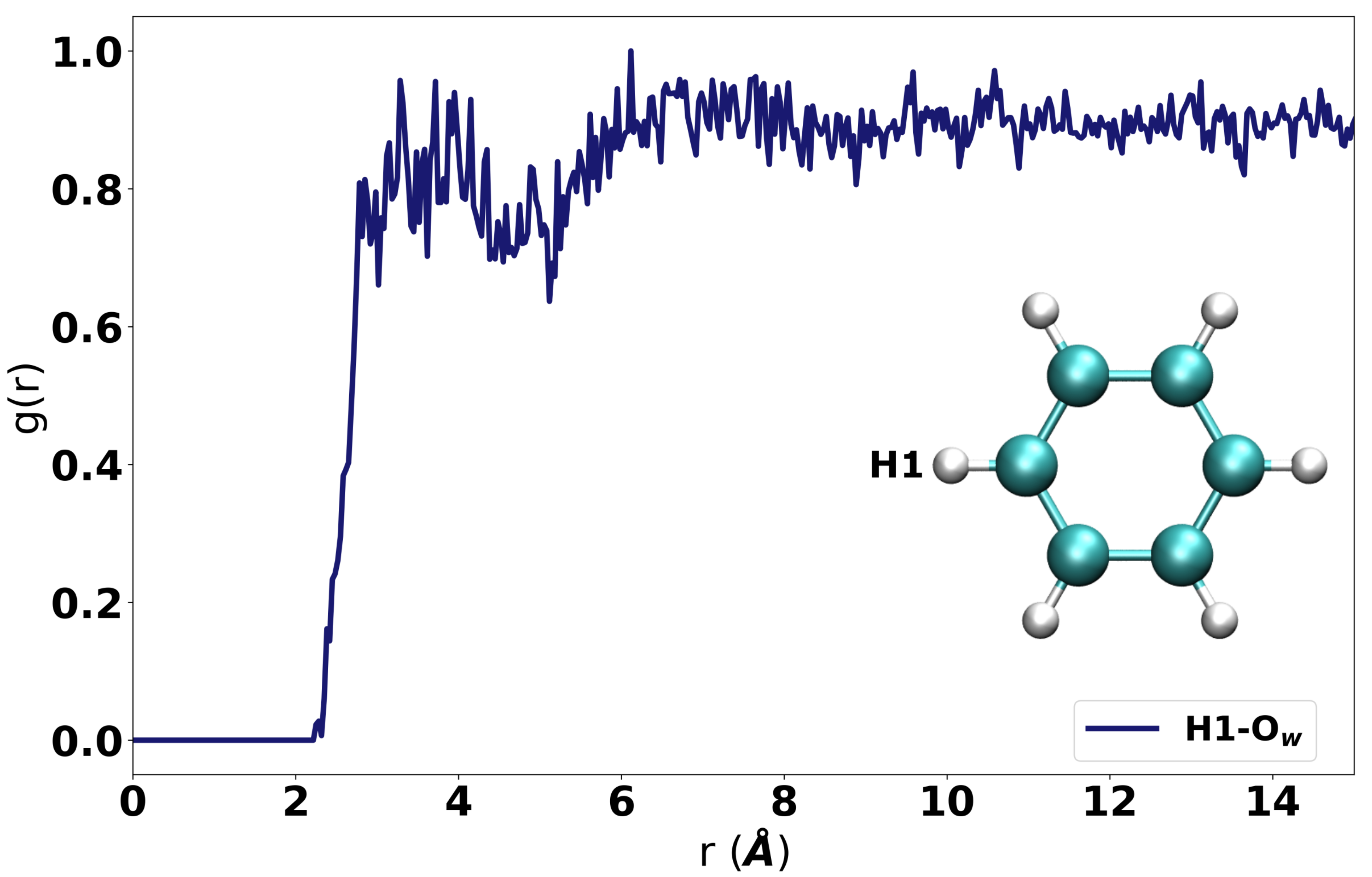}
    \caption{RDF between the benzene hydrogen (H1) and water oxygen atoms.}
    \label{f:rdf_benzene}
    \end{figure}

    \FloatBarrier

\newpage

\section{Active space selection among different frames: MOs overlap} \label{sec:mosoverlap}

To choose the active space consistently among the different spherical snapshots, a modified version of the protocol presented in Ref. \citenum{cardenas2021algorithm} is employed. This procedure aims to select the set of active orbitals manually in a limited number of reference snapshots and then automatically obtain the active space for the rest of the frames (target frames). The idea is to identify the same MOs manually selected in the reference frames in each of the target frames. These orbitals are chosen as the ones that share the maximum overlap with the MOs of the reference frames.
 
 For each target frame, the procedure consists of the following steps: 
 \begin{enumerate}
     \item align the structures of the reference and target frames by computing the optimal rotation matrix using the Kabsch algorithm\cite{lawrence2019purely};
     \item rotate the MOs accordingly, by using the optimal rotation matrix;
     \item  compute the overlap matrix;
     \item select the MOs with the maximum overlap, in absolute value, with the reference MOs.
 \end{enumerate} 

According to this procedure, it is only necessary to inspect the active space for frames where the MOs do not exactly match the ones selected in the reference frames (e.g., when the absolute value of the overlap is less than 0.7).

\section{Non-Orthogonality of SS-CASSCF/FQ Wavefunctions} \label{sec:nonorth}

The impact of non-orthogonality on excitation energies can be estimated through L\"owdin symmetric orthogonalization, which provides the closest orthogonal set to the original states \cite{lowdin1970nonorthogonality,aiken1980lowdin}.
Within this method, the vector that contains the orthogonalized wavefunctions $\boldsymbol{\ket{\Psi'}}$ and the vector that contains the original wavefunctions $\boldsymbol{\ket{\Psi}}$ are related by the following equation:
\begin{equation}
    \boldsymbol{\ket{\Psi'}} = \boldsymbol{S}^{-1/2} \boldsymbol{\ket{\Psi}}
\end{equation}
where $\boldsymbol{S}$ is the overlap matrix between the states:
\begin{equation}
    \boldsymbol{S} = \begin{pmatrix}
\,\,1\, & \,s\,\, \\
\,\,s\, & \,1 \,\,
\end{pmatrix}
\end{equation}
If $s \ll 1$, $\boldsymbol{S}^{-1/2}$ can be approximated as:
\begin{equation}
   \, \boldsymbol{S}^{-1/2} \approx \begin{pmatrix}
1 & -\frac{s}{2} \, \\
-\frac{s}{2} & 1 
\end{pmatrix}
\end{equation}
With this approximation, the orthogonalized states are:
\begin{equation}
\begin{pmatrix}
\Psi'_{GS} \\
\Psi'_{ES} 
\end{pmatrix} = \begin{pmatrix}
1 & -\frac{s}{2} \, \\
\, -\frac{s}{2} & 1 
\end{pmatrix}
\begin{pmatrix}
\Psi_{GS} \\
\Psi_{ES} 
\end{pmatrix}
\end{equation}
Considering the orthogonalized states, the new excitation energy of the system is:
\begin{equation}
\begin{split}
     E'_{ES} - E'_{GS} = & \braket{\Psi'_{ES} | \hat{H} | \Psi'_{ES}} -  \braket{\Psi'_{GS} | \hat{H} | \Psi'_{GS}}  \\ 
    \approx & E_{ES} - E_{GS} - \frac{s^2}{4} (E_{ES} - E_{GS}) 
\end{split}
\end{equation}
where $E_{ES}$ and $E_{GS}$ are the original energies of the ES and GS, respectively. In this equation, an additional approximation has been made, assuming that the GS and ES Hamiltonians are equal. The correction to the excitation energy depends on the square of the overlap term $s$. Selecting a random snapshot of benzene, the excitation energy of the initial states is $E_{ES} - E_{GS} \approx 0.19$ a.u. and the overlap term is $s=-1.7\cdot 10^{-3}$ a.u.. This means that the correction due to the non-orthogonality of the states is within the order of $10^{-7}$ a.u., therefore it marginally affects the value of the excitation energy. Analogously, considering a random snapshot of phenol in aqueous solution, the excitation energy of the initial states is $E_{ES} - E_{GS} \approx 0.18$ a.u. and the overlap between the two wavefunctions is $s=1.0\cdot 10^{-3}$ a.u.. The correction for the excitation energy is therefore of the order of $10^{-8}$ and also in this case does not substantially affect the value of the excitation energy.

 \newpage
 \section{Convergence tests}

\begin{figure}[H]
    \centering
    \includegraphics[width=0.8\linewidth]{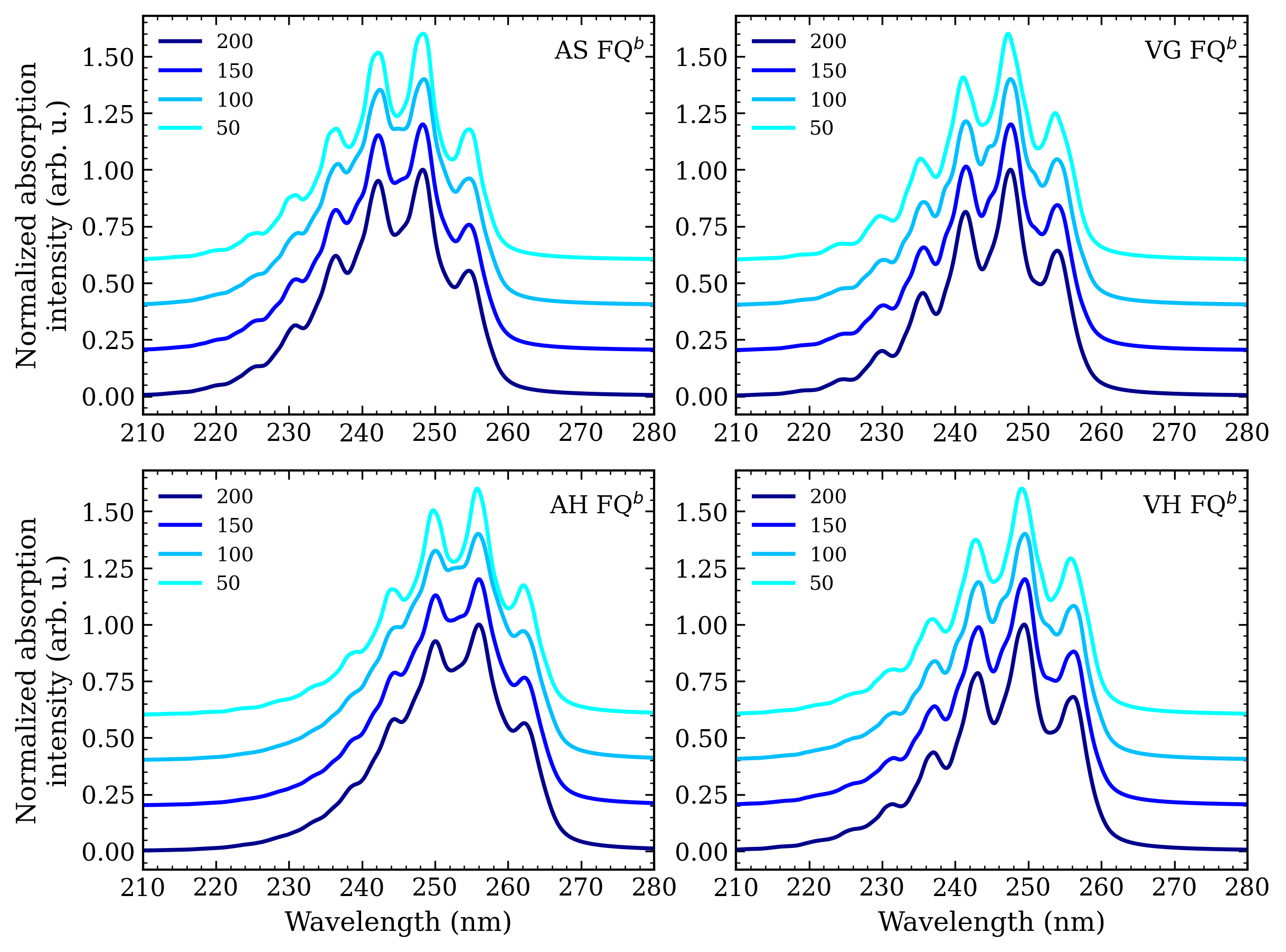}
    \caption{CASSCF(6,6)/FQ$^b$ vibronic spectra of benzene in aqueous solution computed using different harmonic approximations (AS, VG, AH, VH). Each subplot contains 4 stacked spectra obtained by averaging the results of a different number of frames: 50, 100, 150, and 200. Averaged spectra are convoluted with a Lorentzian function with FWHM 0.04 eV.}
    \label{fig:benz_giov_conv}
\end{figure}

\begin{figure}[H]
    \centering
    \includegraphics[width=0.8\linewidth]{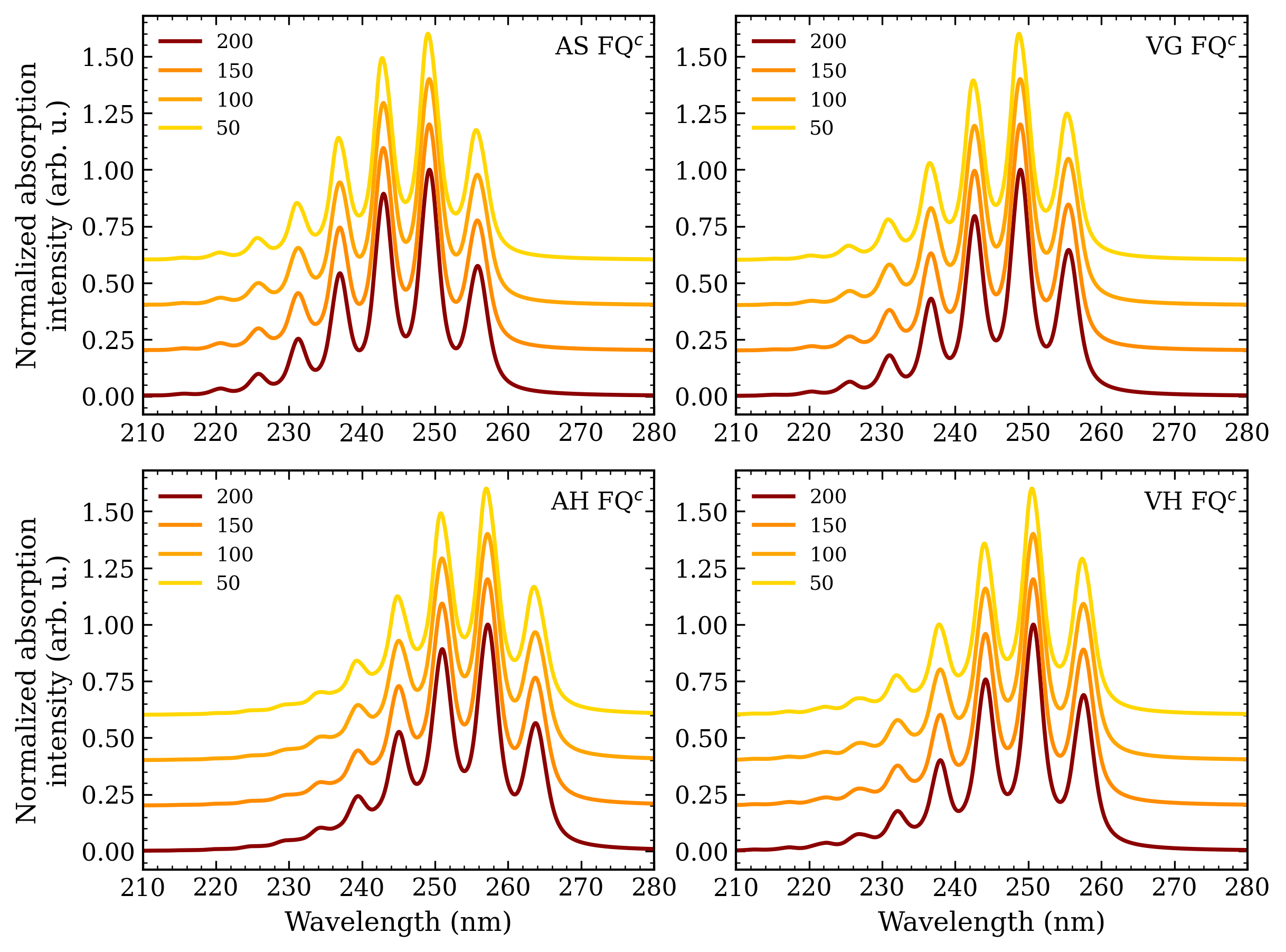}
    \caption{CASSCF(6,6)/FQ$^c$ vibronic spectra of benzene in aqueous solution computed using different harmonic approximations (AS, VG, AH, VH). Each subplot contains 4 stacked spectra obtained by averaging the results of a different number of frames: 50, 100, 150, and 200. Averaged spectra are convoluted with a Lorentzian function with FWHM 0.04 eV.}
    \label{fig:benz_rick_conv}
\end{figure}

\begin{figure}
    \centering
    \includegraphics[width=0.8\linewidth]{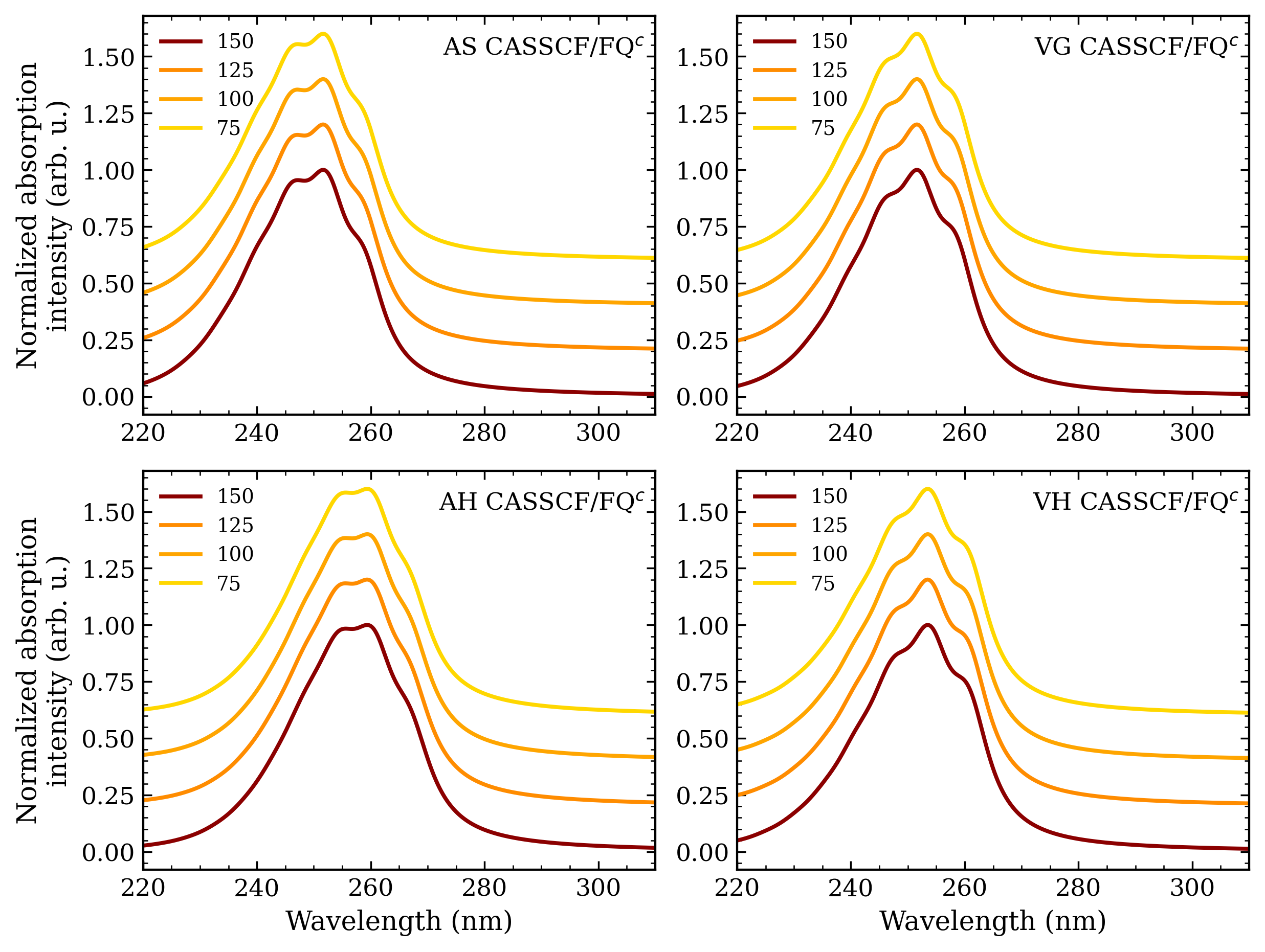}
    \caption{CASSCF(8,7)/FQ$^c$ vibronic spectra of phenol in aqueous solution computed using different harmonic approximations (AS, VG, AH, VH). Each subplot contains 4 stacked spectra obtained by averaging the results of a different number of frames: 75, 100, 125, and 150. Averaged spectra are convoluted with a Lorentzian function with FWHM 0.15 eV.}
    \label{fig:phenol_conv}
\end{figure}

\clearpage

\bibliography{bibliografia_abstract_tesi}